\documentclass[aps,showpacs,twocolumn,eqsecnum,floatfix,amsmath,amssymb]{revtex4}

\newcommand\p{\partial_r}

\usepackage{amsmath,amsfonts,amssymb,amsthm}
\usepackage{epsfig}
\usepackage{graphicx}
\usepackage{latexsym}
\usepackage{bm,latexsym}
\usepackage{mathrsfs}
\usepackage{psfrag}
\usepackage{color}
\begin{document}

%%%%%%%%%%%%%%%%%
%%%   TITLE   %%%
%%%%%%%%%%%%%%%%%

\title{Induced scalarization in boson stars and scalar gravitational radiation}

\author{Milton Ruiz$^1$}
\email{milton.ruiz@uib.es}

\author{Juan Carlos Degollado$^2$}
\email{jdaza@astro.unam.mx}

\author{Miguel Alcubierre$^3$}
\email{malcubi@nucleares.unam.mx}

\author{Dar\'\i o N\'u\~nez$^{3}$}
\email{nunez@nucleares.unam.mx}

\author{Marcelo Salgado$^3$}
\email{marcelo@nucleares.unam.mx}

\affiliation{$^1$Departament de F{\'\i}sica, Universitat de les Illes Balears,
Cra. Valldemossa Km. 7.5, Palma de Mallorca, E-07122, Spain \\
$^2$ Instituto de Astronom\'{\i}a, Universidad Nacional Aut\'onoma de M\'exico, 
Circuito Exterior C.U., A.P. 70-264, M\'exico D.F. 04510, M\'exico \\
$^3$Instituto de Ciencias Nucleares, Universidad Nacional
Aut\'onoma de M\'exico, Circuito Exterior C.U., A.P. 70-543,
M\'exico D.F. 04510, M\'exico}

%%%%%%%%%%%%%%%%
%%%   DATE   %%%
%%%%%%%%%%%%%%%%

\date{\today}

%%%%%%%%%%%%%%%%%%%%
%%%   ABSTRACT   %%%
%%%%%%%%%%%%%%%%%%%%

\begin{abstract}
The dynamical evolution of boson stars in scalar-tensor theories of
gravity is considered in the physical (Jordan) frame.  We focus on the
study of spontaneous and induced scalarization, for which we take as
initial data configurations on the well-known S-branch of a single
boson star in general relativity.  We show that during the
scalarization process a strong emission of scalar radiation occurs.
The new stable configurations (S-branch) of a single boson star within
a particular scalar-tensor theory are also presented.
\end{abstract}

%%%%%%%%%%%%%%%%
%%%   PACS   %%%
%%%%%%%%%%%%%%%%


%%%%%%%%%%%%%%%%%%%%%
%%%   MAKETITLE   %%%
%%%%%%%%%%%%%%%%%%%%%

\maketitle

%%%%%%%%%%%%%%%%%%%%%%%%
%%%   INTRODUCTION   %%%
%%%%%%%%%%%%%%%%%%%%%%%%

\section{Introduction}
\label{sec:introduction}

Scalar-tensor theories (STTs) are alternative theories of gravitation
where a spin$-0$ degree of freedom $\phi$ can accompany the usual
tensor spin$-2$ modes (see Ref.~\cite{Damour92} for a review). There
are two mathematical representations of the STT: 1) The physical frame
(also known as the ``Jordan'' frame), where test particles follow
geodesics of spacetime and the scalar field $\phi$ couples
non-minimally to the curvature: 2) The Einstein frame, obtained by a
conformal transformation of the metric, where the scalar field couples
minimally to the curvature and non-minimally to the matter
fields~\cite{Fujii2003}.

STTs are perhaps the simplest, well motivated and most competitive
theories of gravitation after General Relativity (GR), the most
prominent example being the well-known Jordan-Brans-Dicke
theory~\cite{Jordan55,Brans61}.  Intuitively, STTs can be seen as
theories with a varying effective gravitational ``constant''.
Although so far there is not observational evidence that such scalar
gravitational field exists, one can use the experimental and
observational tests of GR to put limits on its existence and its
possible interactions~\cite{Will:2005va}. Using data from the binary
pulsar, for instance, it is possible to put limits on some classes of
STTs which restrict the form of the non-minimal coupling (NMC) to the
curvature. Nevertheless, these bounds still allow a NMC constant of
order unity~\cite{Damour96}.

Despite the fact that STTs were proposed several decades ago, it has
only been recently that several phenomena associated with them, and
with no counterpart in GR, have been analyzed. For instance, in the
cosmological context, STTs have been proposed as alternatives to the
cosmological constant in order to explain the accelerated expansion of
the
Universe~\cite{Perrotta1999,Boisseau2000,Amendola2001,Riazuelo2002,Salgado02,Schimd2005}.

In the astrophysical scenario, Damour and
Esposito-Far\`ese~\cite{Damour93,Damour96} discovered that neutron
star models within STTs may undergo a {\em phase transition}\/ that
consists on the appearance of a non-trivial configuration of the
scalar field $\phi$ in the absence of sources and with vanishing
asymptotic value. This phenomenon has been named {\em spontaneous
  scalarization}\/ (SS) due to its similarities with the spontaneous
magnetization of ferromagnets at low temperatures.  The stability
analysis for the transition to SS was first performed by
Harada~\cite{Harada97,Harada98}.  It is now understood that SS arises
under certain conditions where the appearance of a non-trivial scalar
field gives rise to a stationary configuration that minimizes the energy of the
star with fixed baryon number.

Further analysis have confirmed that the SS phenomenon takes place in
neutron stars independently of the equation of state (EOS) used to
describe the nuclear matter~\cite{Novak98b,Damour98,Salgado98}. In
boson stars this phenomenon was first studied by
Whinnett~\cite{Whinnett00}, who constructed stationary scalarized
configurations with a self-interaction potential for the scalar field.
More recently, the dynamic transition to SS was analyzed by us in the
Jordan frame and without self-interaction~\cite{Alcubierre:2010ea}.
One important feature of this phenomenon is that it can occur even
when the parameters of the theory satisfy the stringent bounds imposed
by the Solar System experiments, notably, when the Brans-Dicke
parameter is chosen to be arbitrarily large.

The SS phenomenon is accompanied by the ``sudden'' appearance of a new
global quantity termed {\em scalar charge}, where by sudden we mean
that the derivative of this charge with respect to the central energy
density at the critical point is infinite. The scalar charge is the
analogous of the magnetization of ferromagnets mentioned above.
Moreover, just as in the neutron star case, in boson stars one also
finds that beyond a certain critical central energy-density, the
stationary configurations that are energetically preferred are those
where the SS ensues.

A phenomenon similar to SS, but that occurs when a background scalar
field is present, is called {\em induced scalarization}\/ (IS).  It
corresponds to the case where the scalar field does not vanish
asymptotically. In this situation the scalar charge does not exhibit a
discontinuous ``jump'' as the object becomes more compact, and the
transition to scalarization is smoothed out by the presence of the
background field.

Another important feature of STTs is the prediction of scalar
gravitational waves.  While GR predicts only quadrupole gravitational
radiation in the ``far zone'', STTs predict the existence of monopolar
gravitational waves that can be emitted even in the case of spherical
symmetry~\cite{Maggiore2000}.  The new polarization of this scalar
mode is of {\em breathing type}\/ since it affects all directions
isotropically~\cite{Maggiore2000}.

A simple scenario where such scalar gravitational waves might be
produced is precisely during the scalarization process of a spherical
compact object. The amplitude of such waves (in the linear
approximation) is linked directly with both the form of the NMC and
the asymptotic value of the scalar field.  For instance, when this
asymptotic value vanishes, it turns out that certain classes of STTs do
not lead to the emission of scalar gravitational waves. This implies
in particular that in such classes of STTs the SS phenomenon does not
produce monopolar waves, rather it is only in the IS scenario that
such theories can lead to an emission of scalar gravitational
radiation.  Since we will be working with one such class of STTs, it is
then particularly important to make a clear distinction between the SS
and the IS phenomena.

The first dynamical analysis of the scalarization phenomenon in
neutron stars was made by Novak~\cite{Novak98b}.  He did not only
confirm the dynamical transition to the scalarization state, but also
the emission of scalar gravitational waves. Moreover, he also studied
the emission of scalar gravitational radiation when a scalarized star
collapses into a black hole~\cite{Novak98a}.

Recent studies of neutron star oscillations within STTs have shown
that, in addition to the emission of scalar gravitational waves, the
quadrupole gravitational radiation is also disturbed as compared to
the corresponding signals in GR~\cite{Sotani05}.  Therefore, even if
the detection channels for scalar gravitational waves are ``switched
off'', the detection of gravitational waves of spin-2 coming from
these sources might still validate STTs or put even more stringent
bounds on their parameters. Of course, the direct detection of scalar
gravitational waves (or the absence thereof) would also help to
discriminate between several alternative theories.

In this work, we present a systematic study of the phenomenon of IS in
the spacetime of a single boson star, without a self-interaction
potential for the non-minimally coupled scalar field. We choose to
work directly in the physical Jordan frame where the physics is better
understood. In addition, we study some properties of the gravitational
scalar waves such as their magnitude and frequency.  All the numerical
evolutions are performed in spherical symmetry using a 3+1 formalism
of STTs as presented in~\cite{Salgado06}. However, instead of evolving
the geometry with the Arnowitt-Deser-Misner (ADM) equations, we use a
strongly hyperbolic version similar to the
Baumgarte-Shapiro-Shibata-Nakamura (BSSN)
formulation~\cite{Shibata95,Baumgarte:1998te} but adapted to the
STT~\cite{Salgado:2008xh}. Using this 3+1 system the initial value
problem of STTs in the Jordan frame turns to be
well-posed. Nevertheless, at the moment we only have numerical
evidence to show that the initial {\em boundary}\/ value problem
(IBVP) is also well-posed. The analysis of the continuum IBVP for this
system will be presented elsewhere.

This paper is organized as follows. Section~\ref{sec:STT} introduces
the STTs and discusses briefly some properties associated with the
Jordan frame. The relevant 3+1 equations of~\cite{Salgado06} are also
presented.  For completeness, and for the benefit of the reader, in
Section~\ref{spontaneos-scalar} we discuss the heuristic analysis
performed by Damour and Esposito-Far\`ese in~\cite{Damour93}, which
allows one to understand the scalarization phenomenon on simple
grounds.  Section~\ref{sec:bosons} contains our boson star model. In
Section~\ref{sec:waves} we describe the scalar waves predicted in
STTs. Section~\ref{sec:numest} summarizes the setup used in the
numerical code.  We present the results of our numerical simulation in
Section~\ref{sec:numres}, and we conclude in
Section~\ref{sec:discussion}.

We also present in Appendix~\ref{sec:Charac-BSSN} the characteristic
decomposition of the spherically symmetric equations used in this
paper, which allow us to conclude that our system is strongly
hyperbolic and, therefore, that the Cauchy problem is well-posed for
the spherically symmetric case. In Appendix~\ref{sec:CPBCs} we present
some numerical evidence that indicates that corresponding IBVP is also
well-posed.

%%%%%%%%%%%%%%%%%%%%%%%%
%%%   STT THEORIES   %%%
%%%%%%%%%%%%%%%%%%%%%%%%

\section{Scalar Tensor Theories of gravity}
\label{sec:STT}

\subsection{Field equations} 
\label{sec:fieldequations}

The STTs of gravitation are one of the simplest and most analyzed
alternative theories of gravity.  These alternative theories were
introduced by Jordan during the decade of the fifties~\cite{Jordan55},
and then reanalyzed by Brans and Dicke later~\cite{Brans61}. The
general action for STTs in the Jordan frame, where gravity is coupled
non-minimally to a single scalar field $\phi$, is given by
\begin{eqnarray}
S[g_{ab},\phi,\psi] &=&
\int \left\{\frac{F(\phi)}{16\pi G_0} R
- \frac{1}{2}(\nabla \phi)^2 - V(\phi)\right\} \sqrt{-g} \: d^4x
\nonumber\\
&+& S_{\rm matt}[g_{ab}, \psi] \; ,
\label{eq:jordan}
\end{eqnarray}
where $\psi$ represents all the matter fields, {\em i.e.} fields other
than $\phi$, $G_0$ is the usual gravitational constant, $F(\phi)$ is
some NMC function to be specified later, and $V(\phi)$ represents a
potential for $\phi$ (we use units such that $c=1$). In fact, in all
the numerical analysis considered here, we will not consider the
potential $V(\phi)$.  However, for completeness it will be included in
the field equations displayed below. Notice that one can identify the
``effective'' gravitational constant as the coefficient $G_{\rm
  eff}(\phi)=G_0/F(\phi)$ that appears in the above
action~\footnote{Here and in what follows, Latin indices from the
  first part of the alphabet $a,b,c,\cdots$ will denote
  $4$-dimensional quantities, while Latin indices from the middle of
  the alphabet $i,j,k,\cdots$ will denote $3$-dimensional spatial
  quantities.}.

From the above action, one finds the following field equations:
\begin{align}
R_{ab}&-\frac{1}{2}g_{ab}R = 8 \pi G_0 T_{ab} \; , 
\label{eq:Einst} \\
\Box \phi &+ \frac{1}{2}f^\prime R = V^\prime\,,
\label{eq:KGo}
\end{align}
where a prime indicates $\partial_\phi$,
$\Box:=g^{ab}\nabla_a\nabla_b$ is the standard covariant d'Alambertian
operator, and
\begin{align}  
T_{ab} &:= \frac{G_{{\rm eff}}}{G_0} \left( \rule{0mm}{0.5cm} T_{ab}^f
+ T_{ab}^{\phi} + T_{ab}^{{\rm matt}}\right) \; , 
\label{eq:effTmunu} \\
T_{ab}^f &:= \nabla_a \left( f^\prime 
\nabla_b\phi\right) - g_{ab}\nabla_c \left(f^\prime 
\nabla^c \phi\right) \; ,
\label{eq:TabF} \\
T_{ab}^{\phi} &:=  (\nabla_a \phi)(\nabla_b \phi) - g_{ab}
\left[ \frac{1}{2}(\nabla \phi)^2 + V(\phi) \right ] \; , \quad
\label{eq:Tabphi}
\end{align}
with $T_{ab}^{\rm matt}$ the stress-energy tensor of all matter fields
other than $\phi$, and where we have defined
\begin{equation} 
f := \frac{F}{8\,\pi\,G_0} \; , \qquad
G_{{\rm eff}} := \frac{1}{8\,\pi\,f} \; .
\label{eq:Geff}
\end{equation}
Notice that equation~\eqref{eq:Einst} implies that the Ricci scalar
can be expressed in terms of the trace of the energy-momentum
tensor~\eqref{eq:effTmunu}.  Therefore, Eq.~\eqref{eq:KGo} can be
rewritten in the form
\begin{equation}
\Box\phi= \frac{2\,f V^\prime - 4\,f^\prime V -f^\prime
\left( 1 +  3f^{\prime\prime} 
\right)(\nabla \phi)^2 + f^\prime T_{{\rm matt}} }
{2\,f\left(1 + 3{f^\prime}^2/2f \right) } \; ,
\label{eq:KG}
\end{equation}
with $T_{{\rm matt}}$ the trace of $T_{ab}^{{\rm matt}}$.  On the
other hand, the Bianchi identities directly imply
\begin{equation}
\nabla _{c }T^{c a } = 0\,.
\end{equation}
Nevertheless, the use of the field equations leads to the conservation
of the energy-momentum tensor of the matter alone
\begin{equation}
\nabla _{c }T_{{\rm matt}}^{c a } = 0 \,,
\end{equation}
which implies the fulfillment of the (weak) equivalence principle,
{\it i.e.} test particles follow geodesics of the metric $g_{ab}$.

%%%%%%%%%%%%%%%%%%%%%%%%%%%%%
%%%   3+1 DECOMPOSITION   %%%
%%%%%%%%%%%%%%%%%%%%%%%%%%%%%

\subsection{3+1 decomposition} 
\label{sec:3+1}

In order to recast the previous field equations as a Cauchy initial
value problem~\cite{York79,Gourgoulhon2012}, we first rewrite the
four-dimensional metric in 3+1 form as
\begin{equation}
ds^2 = - \left( \alpha^2 - \beta^i \beta_i \right) dt^2
+ 2\,\beta_i dx^{i} dt + \gamma_{ij} dx^{i} dx^{j} \, ,
\label{metric}
\end{equation}
with $\alpha$ the lapse function, $\beta^i$ the shift vector, and
$\gamma_{ij}$ the 3-metric induced on the spatial hypersurfaces.

We perform a $3+1$ decomposition of equations \eqref{eq:Einst} and
\eqref{eq:KGo}, using the normal timelike vector $n^a$ to the
spacelike hypersurfaces $\Sigma_t$, and the projection operator
\mbox{$P^{a}{}_{b} = \delta^{a}{}_{b} + n^a\,n_b$}.  In order to do
so, we first define the first order variables:
\begin{align}
Q_i & :=  D_i \phi = P^{k}_{i}\nabla_k\phi  \; ,
\label{eq:Q} \\
\Pi & := n^{a}\nabla_a\phi = \frac{1}{\alpha} \frac{d\phi}{dt}\; , 
\label{eq:Pidef}
\end{align}
where $D_i$ is the covariant derivative compatible with the 3-metric
$\gamma_{ij}$, and \mbox{$d/dt := \partial_t - \mathcal {L}_{\beta}$},
with $\mathcal {L}_{\beta}$ the Lie derivative along the shift vector.
Notice that the relevant components of the quantities computed with
${P^a}_{b}$ are the spatial ones. It is now straightforward to show
that $Q_i$ and $\Pi$ evolve according to~\cite{Salgado06},
\begin{eqnarray}
\frac{dQ_i}{dt} &=& D_i(\alpha\,\Pi) \; ,
\label{eq:EvQ} \\
\frac{d\Pi}{dt} &=&
\alpha \left[ \Pi K + Q^l D_l ({\rm ln} \alpha) + D_l Q^l \right]
\nonumber \\
&-& \frac{\alpha}{2 f \left( 1 + \frac{3{f^\prime}^2}{2f} \right)} 
\left[ \rule{0mm}{4mm} 2 f V^\prime - 4 f^\prime V 
\right. \nonumber \\
&-& \left. \rule{0mm}{4mm} f^\prime \left( 1 +  3 f^{\prime\prime} \right)
 \left( Q^2 - \Pi^2 \right) + f^\prime T_{\rm matt} \right] . \hspace{3mm}
\label{eq:evPi1}
\end{eqnarray}

We define the energy density \mbox{$\rho := n^a n^b T_{ab}$}, momentum
density $J_a := - P^{b}{}_{a} n^c T_{bc}$ and a stress tensor
\mbox{$S_{ab} := P^{c}{}_{a} P^{d}{}_{b}
  T_{cd}$}. From~\eqref{eq:effTmunu} we find
\begin{eqnarray}
\rho &=& \frac{G_{\rm eff}}{G_0} \left( \rho^f + \rho^\phi
+ \rho^{\rm matt} \right) \; ,
\label{eq:ESTT} \\
J_i &=& \frac{G_{\rm eff}}{G_0} \left( J^f_i + J^\phi_i
+ J^{\rm matt}_i\right) \; , \\
S_{ij} &=& \frac{G_{\rm eff}}{G_0} \left( S^f_{ij} + S^\phi_{ij}
+ S^{\rm matt}_{ij}\right) \; . 
\label{eq:SabSTT}
\end{eqnarray}

Using now Eqs.~\eqref{eq:Geff} and \eqref{eq:KG}, one can show
that~\cite{Salgado06}
\begin{eqnarray}
\rho &=& \frac{1}{8\pi G_0 f} \left[ f^\prime \left( D_k Q^k + K \Pi \right)
+ \frac{\Pi^2}{2} \right. \nonumber \\
&+& \left. \frac{Q^2}{2} \left( 1 + 2f^{\prime\prime}\right) 
+ V(\phi) + \rho_{\rm matt} \rule{0mm}{5mm} \right] \; ,
\label{totE} \\
J_i &=& \frac{1}{8\pi G_0 f} \left[ \rule{0mm}{5mm} 
- f^\prime \left( \rule{0mm}{0.4cm} K_i^k Q_k + D_i \Pi \right)
\right. \nonumber \\
&-& \left. \Pi Q_i \left( \rule{0mm}{0.4cm} 1 + f^{\prime\prime} \right) 
+ J^{\rm matt}_i \rule{0mm}{5mm} \right] \; ,
\label{Ji} \\
S_{ij} &=& \frac{1}{8\pi G_0 f} \left\{ \rule{0mm}{6mm}
Q_i Q_j \left( 1 + f^{\prime\prime} \right)  
+  f^\prime \left( D_i Q_j + \Pi K_{ij} \right) \right. \nonumber \\ 
&+& \left.\frac{\gamma_{ij}}{\left(1 + \frac{3{f^\prime}^2}{2f}\right)}
\left[ \frac{1}{2} \left( \rule{0mm}{0.4cm} Q^2-\Pi^2 \right)
\left( 1 + \frac{{f^\prime}^2}{2f}
+ 2 f^{\prime\prime}\right) \right. \right. \nonumber \\
&+& \left. \left.
V\left(1- \frac{{f^\prime}^2}{2f}\right) + f^\prime V^\prime + 
\frac{{f^\prime}^2}{2f} \left( \rule{0mm}{4mm} S_{\rm matt}
- \rho_{\rm matt} \right) \right] \right. \nonumber \\
&+& \left. S^{\rm matt}_{ij} \rule{0mm}{6mm} \right\} \; .
\label{Sab}
\end{eqnarray}
where we have defined $Q^2:= Q^l Q_l$. 

Notice that the $3+1$ decomposition of~\eqref{eq:Einst} are just the
usual ADM equations given by
\begin{eqnarray}
\frac{d\gamma_{ij}}{dt} &=& - 2\, \alpha\, K_{ij} \; , 
\label{eq:gammadot} \\
\frac{dK_{ij}}{dt} &=& - \nabla_i\nabla_j \alpha
+ \alpha \left[ R_{ij} + K\, K_{ij} - 2 K_{il} {K^l}_j \right]
\nonumber \\
&+& 4 \pi G_0 \alpha \left[ \gamma_{ij} \left( S - \rho \right)
- 2 S_{ij} \right] \; ,
\label{eq:Kdot}
\end{eqnarray}
where $R_{ij}$ is the 3-dimensional Ricci tensor associated with the
spatial metric $\gamma_{ij}$, and the effective matter terms are given
by Eqs.~\eqref{totE}$-$\eqref{Sab}.  The Hamiltonian and momentum
constraints take the form
\begin{eqnarray}
H &:=& \frac{1}{2} \left( R + K^2 - K_{ij}\,K^{ij} \right)
- 8 \pi G_0 \rho = 0 \; , 
\label{eq:hamiltonian} \\
M^i &:=& D_l \left( K^{il} - \gamma^{il} K \right)
- 8 \pi G_0 J^i = 0 \; . 
\label{eq:momentum}
\end{eqnarray}

Formally, one should also consider the constraint $D_{[i}Q_{j]}=0$,
which corresponds to the integrability condition $\partial_{ij}^2\phi=
\partial_{ji}^2\phi$.  The above system of evolution equations has to
be completed with appropriate evolution equations for the gauge
variables. This issue is considered below.

%%%%%%%%%%%%%%%%%%%%%%%%
%%%   GAUGE CHOICE   %%%
%%%%%%%%%%%%%%%%%%%%%%%%

\subsection{Gauge choice}
\label{sec:gauge}

To obtain a closed evolution system, one has to impose gauge
conditions for the time variable $t$ and for the spatial coordinates
$x^i$. Following~\cite{Salgado:2008xh}, we will consider a modified
Bona-Masso (MBM) time slicing condition~\cite{Bona94b}. In local
coordinates adapted to the $3+1$ foliation $x^a=(t,x^i)$, this slicing
condition is given by
\begin{equation} 
\frac{d \alpha}{dt} = - \alpha^2 f_{BM}(\alpha)
\left[ K - \frac{\Theta}{f_{BM}(\alpha)} \: \frac{f^{\prime}}{f} \: \Pi\right] \; ,
\label{STTBMlapse}
\end{equation}
with $f_{BM}(\alpha)>0$ the usual Bona-Masso gauge function and
$\Theta$ a free parameter.

The specific choices $\Theta=f_{BM}= 1$ correspond to a modified
harmonic slicing condition (termed ``pseudo-harmonic''
in~\cite{Salgado06,Salgado:2008xh}), which was specially useful for
the second order hyperbolicity analysis performed
in~\cite{Salgado06}. On the other hand, with $\Theta= 0$ one recovers
the usual Bona-Masso (BM) slicing condition. However, it has been
shown that taking $\Theta= 0$ does not result in a strongly hyperbolic
formulation of STTs in the Jordan frame~\cite{Salgado:2008xh}. For this
reason, in all the simulations presented here we have used the
pseudo-harmonic slice with $\Theta=1$.

Concerning the propagation of the spatial coordinates, we will
consider the shift vector as an {\em a priori}\/ known function of the
coordinates.  In particular, in all our evolutions it is set to zero.
However, in the future, it would be interesting to investigate some
``live'' shift conditions and their effects in phenomena involving
STTs.

%%%%%%%%%%%%%%%%%%%%%%%%%%%%%%%%%%%%%
%%%   SPONTANEOUS SCALARIZATION   %%%
%%%%%%%%%%%%%%%%%%%%%%%%%%%%%%%%%%%%%

\section{Scalarization}
\label{spontaneos-scalar}

The STTs of gravity induce strong field effects which, for instance,
produce important deviations from GR in stellar models.  As mentioned
before, one such effect is the SS phenomenon which is similar to the
spontaneous magnetization in ferromagnetic materials at low
temperatures.  In the following, we will use this analogy in order to
understand the SS phenomenon.
 
When a ferromagnet is exposed to an external magnetic field, the
individual spins of its constituents align with the field, giving rise
to a permanent magnetization which remains even after the external
field is switched off. Moreover, the ferromagnets have the property
that, below the Curie temperature, a magnetization appears
``spontaneously'' even in the absence of an external magnetic
field. In the STTs, on the other hand, a nontrivial configuration of a
scalar field may spontaneously appear during the evolution of a
compact object in the absence of external sources, {\it i.e.} without
a potential $V(\phi)$.  One can then identify the external magnetic
field with a background (cosmological) scalar field and the
temperature with the inverse of the central energy-density
$\rho_c^{\rm matt}$ of the matter content or, equivalently, with the
inverse of total baryon mass (in the case of neutron stars).  The role
of the magnetization is played by a new global quantity called the
{\it scalar charge}\/ $Q_{scal}$, which will be defined below and
which corresponds to the coefficient of $\phi(r) \sim Q_{scal}/r$ in
the asymptotic region.  This means that beyond a certain critical
density or critical baryon mass, the transition to the spontaneous
scalarization ensues.  In this case $\partial Q_{scal}/\partial
\rho_c^{\rm matt}$ is infinite at the critical energy-density.  This
transition can be smoothed by the presence of a non-zero background
scalar field $\phi_0$.

In practice, when the phenomenon is analyzed in static configurations,
the value $\phi_0$ is usually fixed by a shooting
method~\cite{Press86,Salgado98}.  It is important to emphasize the
relevance of not adding sources to the scalar field equation. Indeed,
a non-zero potential can make the scalar field decrease so fast that
the scalar charge might in fact vanish at infinity.

An interesting analytical toy model to understand this phenomenon is
the following~\cite{Damour96}.  Consider a static and spherically
symmetric compact object represented by an incompressible fluid
(constant energy-density), whose profile density is given by a step
function. Moreover, assume that the function $F(\phi)$ in the action
for STTs is just a quadratic function of, $F(\phi)= 1 + 8\pi G_0 \xi \phi^2$, and the
potential $V(\phi)$ vanishes. Finally, assume $R\approx
- 8 \,\pi\,G_0\, T_{\rm matt}\approx 8\,\pi\,G_0\,\rho_{\rm matt}$ (as
in GR). This implies, for instance, that the generalized Klein-Gordon (KG)
Eq.~(\ref{eq:KGo}) is linear in $\phi$.  Furthermore, by neglecting
all the gravitational effects in this equation, one ends-up with a
Helmholtz-like equation of the form \mbox{$\Delta \phi + k^2\phi =0$},
where the mass-like term $k^2= 8\,\pi\,G_0\,\xi \,\rho_{\rm matt}$
depends on both the constant energy-density of the matter and the NMC
constant $\xi$. Notice that in this simplified model, the mass term
vanishes beyond the surface of the compact object. Now, for $\xi >0$
the interior regular spherically symmetric solution of the above
Helmholtz equation is given by
\begin{equation}
\phi_{\rm int}(r)=\phi_c \frac{\sin(k\,r)}{k\,r}\,,
\end{equation}
where $\phi_c$ is the scalar field at the origin $r=0$. The exterior
solution is
\begin{equation}
\phi_{\rm ext}(r)= \frac{C}{r} + \phi_0\,,
\end{equation}
where $C$ is an integration constant and $\phi_0$ is the asymptotic
value of the scalar field.

When both solutions are matched continuously at the surface of the
object $r={\cal R}$ [$\rho_{\rm matt}(r\geq {\cal R})=0$], it turns
out that $\phi_c= \phi_0/\cos(k\,{\cal R})$. The explicit form for the
constant $C$ is not relevant for the analysis (but one finds $C\propto
\phi_c {\cal R}$).  Note that if $\phi_0=0$, automatically
$\phi_c=0=C$ and, therefore, $\phi(r)\equiv 0$. In this case, there is
no scalarization. On the other hand, a different situation can happen
if $\cos(k{\cal R})$ vanishes as well when $\phi_0\rightarrow 0$. This
can occur when $k= \pi/(2{\cal R})$. In this case, the solution is
given by
\begin{align}
\label{phiaproxSC1}
\phi_{\rm int}(r)&=\frac{\epsilon\,\sin(\bar{r})}{\bar{r}}\,,\\
\label{phiaproxSC2}
\phi_{\rm ext}(r)&=\frac{\epsilon}{\bar{r}}\,,
\end{align}
where $\epsilon$ is a constant related to the scalar charge $Q_{scal}$
whose numerical value depends on the details of the model, and ${\bar
  r}=r\,\pi/(2\,{\cal R})$.  The above simplified analysis agrees
qualitatively with the full numerical study
~\cite{Damour96,Salgado98,Novak98b}.  Note that if $\xi <0$, the
interior solution is $\phi_{\rm int}(r)=\phi_c \sinh(|k|\,r)/(|k|\,r)$
and $\phi_c= \phi_0/\cosh(|k|\,{\cal R})$. In this case, the scenario
is completely different since the function $\cosh(|k|\,{\cal R})$
never vanishes, and so when $\phi_0\rightarrow 0$ then automatically
$\phi_c\rightarrow 0$ and then the scalar field vanishes everywhere
(no scalarization ensues). It is somehow remarkable that the
spontaneous scalarization phenomenon is associated with a decreasing
effective gravitational constant ({\it i.e.} $G_{\rm eff}< G_0$)
~\cite{Salgado98}.

Another way to understand the existence of these kind of scalarized
configurations is to notice that the presence of a non-trivial scalar
field $\phi(r)$ (within a class of STTs) causes the total energy of the 
stationary configuration to decrease relative to the case where $\phi(r)=0$ for a
fixed baryon mass~\cite{Damour93,Salgado98}.  This can also be
understood on Newtonian grounds by a suitably redefinition of the kind
of energy that has to be minimized when dealing with a theory where an
effective gravitational ``constant'' may vary~\cite{Salgado98}.  The
energetic analysis shows that for large compactness, the energetically
preferred stationary configurations are those with a non-trivial scalar
field. Again, in the ferromagnetic analogy, one appreciates that in
the Landau ansatz, the free energy of the ferromagnet becomes lower in
the presence of magnetization than the energy in the absence of it
when the temperature is below the Curie point. This occurs since below
that temperature the free energy develops a global minimum and a local
maximum (like a Mexican hat potential). The local maximum of the free
energy is located at zero magnetization while the minimum corresponds
to a non-zero magnetization.  Recently, it was also been argued that
the SS phenomenon can be traced back to the quantum fluctuations of
the vacuum state associated with the scalar
field~\cite{Pani2011,Lima2010}.

The scalar charge which characterizes the scalarized configuration is
defined as
\begin{equation}
{Q}_{scal}:= -\lim_{r \rightarrow \infty} \frac{1}{4\pi\,\sqrt{G_0}}
\int_{S} s_a \nabla^a\,\phi\,ds \; ,
\end{equation}
where $s^a$ is the unit outward normal to a topological 2-sphere $S$
embedded in $\Sigma_t$, and $r$ is a radial coordinate that provides
the area of $S$ asymptotically. As it was remarked in the
introduction, when the asymptotic value of the scalar field, $\phi_0$,
is not demanded to vanish but is only accommodated to satisfy the
Solar System bounds, then the scalarization process is {\it induced}\/
by such background (cosmological) field. In such situation the
transition from a small scalar charge to a large one (which depends on
the compactness of the object) is smoothed out and the derivative
$\partial Q_{scal}/\partial \rho_0^{\rm matt}$ is always finite. In
this paper, we will be concerned with this latter situation only, but
as long as $\phi_0$ is small, the difference between the two type of
scalarizations is just a matter of principle. Nevertheless, the
important point for making such a distinction is that while in the SS
case there are no emission of scalar gravitational waves (when
$F'(\phi)_{\phi_0=0}=0$, which is the case for the quadratic function
$F(\phi)$ considered above), for the {\it induced}\/ case where
$\phi_0\neq 0$, one can have a small but non-zero amplitude for the
scalar waves ({\it c.f.}  Sec.~\ref{sec:waves}).

Another important qualitative mathematical aspect that distinguish
both types of scalarizations for a NMC like the quadratic one is the
following.  If one considers Eq.~(\ref{eq:KGo}) in absence of a
potential, it turns out that $\phi=0$ is always a stationary solution
of the equation.  This implies then that $\phi_0\equiv 0$. Therefore,
in order to trigger the transition to a SS case an explicit
scalar-field perturbation is required. An analysis of this sort was
performed by us in~\cite{Alcubierre:2010ea}. However, if one considers
initially a trivial (but non-zero) scalar-field configuration $\phi=
\phi_0=const.$, then this is not a stationary solution of
Eq.~(\ref{eq:KGo}). In such a case {\it a fortiori}\/ the scalar field
will evolve in time without the need of any perturbation. How much of
the initial energy of the star will then be transformed into scalar
energy and scalar radiation leaving behind a highly non trivial
stationary scalar field configuration will depend precisely on the
compactness of the object.  Higher compact objects will radiate more
energy in the form of scalar radiation than lower compact
ones. Therefore, higher compact objects are expected to end up in a
stationary scalarized state with energy lower than the initial one,
the difference being radiated away in scalar-field form.  Clearly, in
order to analyze in detail the transition towards a scalarized state
from a state with a trivial (non-zero) scalar field and its
corresponding emission of scalar radiation, a dynamical evolution is
required.  This is the aim of this paper.

%%%%%%%%%%%%%%%%%%%%%%%
%%%   BOSON STARS   %%%
%%%%%%%%%%%%%%%%%%%%%%%

\section{Boson Stars}
\label{sec:bosons}

Boson stars are equilibrium configurations of a self-gravitating
(condensate) complex scalar field. Their ``hydrostatic'' equilibrium
is maintained by the intrinsic effective pressure of the boson field
due to the uncertainty principle (for a review see
Ref.~\cite{Jetzer92,Liebling:2012fv}), rather than the Pauli exclusion
principle that acts, for instance, in neutron stars. Classically, one
can interpret the equilibrium as a consequence of an effective
pressure associated with the boson field which depends on its
gradients and potential. Since the energy-density and pressure are
parameterized in a certain way by the boson field itself, the relation
between them provides a non-trivial EOS for this kind of matter.

Boson stars are also interpreted as macroscopic boson quantum states
whose associated physical particles are formed by the excitations
around the vacuum expectation value of the scalar field.  The
theoretical existence of such objects were proven first by
Kaup~\cite{Kaup68}, and latter by Ruffini and
Bonazzola~\cite{Ruffini69}, for the ground state solutions of a free
boson field.  Using the uncertainty principle and the definition of
the Schwarzschild radius, it can be shown that the boson stars
considered there have an effective radius $R_{eff}\sim \hbar/m_b\,c$
and a maximum mass of
\begin{equation}
M_{max}\sim \frac{\hbar\,c}{2\,G_0\,m_b}= 0.5\,M_{Pl}^2/m_b\,,
\end{equation}
where $M_{Pl}$ is the Planck mass and $m_b$ is the mass associated
with bosons (for clarity we have restored the speed of light $c$).
Numerical results show that, in fact, this limit is
$M_{max}\thickapprox (2/\pi)\,M_{Pl}^2/m_b$.  Therefore, the resulting
sizes and masses of boson stars would be so small as to be
astrophysically inconsequential. In a more recent paper
\cite{Colpi86}, it was shown that a self-interacting boson field (with
an interaction of the form $\sim\lambda \phi^4$) can give rise to
stable boson stars with
\begin{equation}
M_{max}\sim \lambda^{1/2}\,M_{Pl}^2/m_b \sim {\rm GeV}^2\,
\lambda^{1/2}\,M_\odot/m_b^2\,,
\end{equation}
which is comparable with the Chandrasekhar mass for fermion
stars~\cite{Mielke:2000mh}.  This result was then extended to the
so-called soliton stars \cite{Lee87,Friedberg87}.

Boson star models have been constructed in the past within the
framework of STTs (see {\it e.g.}
\cite{Torres97,Comer97,Torres98a,Torres98b,Whinnett00,Alcubierre:2010ea,Horbatsch:2010hj}
and references therein), although only Whinnett had shown that the
phenomenon of spontaneous scalarization occurs in these objects with
the inclusion of a quartic self-interaction
potential~\cite{Whinnett00}.  Recently, we have found that
self-interactions are not in fact necessary in order to produce
scalarization~\cite{Alcubierre:2010ea}. On the other hand, Torres has
shown in~\cite{Torres97} that, for parameters and boundary conditions
respecting the weak-field limits and the nucleosynthesis bounds, the
masses of boson stars in the STTs framework are comparable with the
ones in GR (for stars in the ground state). Comer~\cite{Comer97}
confirmed the same trend for the case of boson stars in ``excited''
states. Equilibrium and stability properties for these stars in STTs
for different cosmic ages have been analyzed
in~\cite{Torres98a,Torres98b}.

An important aspect of analyzing boson stars in the framework of STTs
is that the transition to a scalarized state might be accompanied by
the emission of (spin-0) scalar gravitational waves (like in neutron
stars).  It is possible that such kind of waves might be detected in
the future if fundamental scalar fields do exist in
nature~\cite{Maggiore2000}.

%%%%%%%%%%%%%%%%%%%%%%%%%%%%%
%%%   LAGRANGIAN FOR BS   %%%
%%%%%%%%%%%%%%%%%%%%%%%%%%%%%

\subsection{The Model}

Boson stars are described by the Lagrangian density of a complex
scalar field
\begin{eqnarray}
\mathcal{L}_{\rm matt} = -\frac{1}{2} g^{ab}\,\nabla_a\psi\nabla_b \psi^* -
V_{\psi}(|\psi|^2)\,,
\label{eq:LKG}
\end{eqnarray}
where $\psi$ is the scalar field, $\psi^*$ its complex conjugate,
$|\psi|^2=\psi~\psi^*$, and $V_{\psi}(|\psi|^2)$ is a potential
depending just on the norm.  Variation of the above Lagrangian with
respect to $\psi$ leads to the KG equation
\begin{equation}
{\Box \psi} = 2\,\psi\, \frac{dV_{\psi}}{d|\psi|^2}\,.
\label{eq:KGbos}
\end{equation}
It is straightforward to show that if the scalar field $\psi$ is real,
the KG equation (\ref{eq:KGbos}) takes the usual form \mbox{${\Box
    \psi} = \partial_{\psi} V_\psi$}.  On the other hand, variation
of~(\ref{eq:LKG}) with respect to the metric $g_{ab}$ leads to the
energy-momentum tensor
\begin{eqnarray}
T_{ab}^{\rm matt} &=& \frac{1}{2} \left[ \rule{0mm}{0.4cm}
(\nabla_a \psi^*)(\nabla_b \psi)
+ (\nabla_b \psi^*)(\nabla_a \psi) \right] \nonumber \\
&& - g_{ab}
\left[ \frac{1}{2}\,|\nabla \psi|^2 + V_{\psi}(|\psi|^2) \right ]\,.
\label{eq:bospot}
\end{eqnarray}
We will consider only the free field case where the potential is given
by
\begin{equation} 
V_{\psi}(|\psi|^2)= \frac{1}{2}\, m^2 \psi^* \psi \,, 
\label{eq:potentialKG}
\end{equation} 
with $m$ a parameter that can be consider as the ``bare mass'' of the
theory (although it has units of inverse length). It is possible to
include more general terms in~(\ref{eq:potentialKG}), such as
$\lambda\,|\psi|^4$ which corresponds to a self-interaction
term~\cite{Colpi86}.

The Lagrangian~(\ref{eq:LKG}) is invariant with respect to a global
phase transformation $\psi\rightarrow e^{\imath\,q} \psi$ (with $q$ a
real constant). The Noether theorem then implies the local
conservation of the boson number $\nabla_a {\cal J}^a=0$, where the
number-density current is given by
\begin{equation}
\mathcal{J}^a =\frac{\imath}{2}\,g^{ab}\,\left[\psi\,\nabla_b\psi^* 
- \psi^*\,\nabla_b \psi\right]\,.
\end{equation}
This means that the total boson number,
\begin{equation}
 {\cal N} =- \int_{\Sigma_t}\sqrt{\gamma}\,\,n_a\, {\cal J}^a\, d^3 x\,,
\end{equation}
is conserved (here $\gamma$ is the determinant of the 3-metric
$\gamma_{ij}$). The total boson mass can be defined by
\begin{equation}
 M_{\rm bos}:= m_b\,{\cal N} \,,
\label{eq:Tbosonmass}
\end{equation}
where $m_b:= 2\,\pi\,\hbar\,m/c$ is the mass of single bosons (again
we have restored the speed of light $c$).

A $3+1$ decomposition of the above energy-momentum tensor gives rise
to the following matter variables
\begin{align}
\rho_{\rm matt} &= \frac{1}{2}\,\left( |\Pi_\psi|^2 + |Q_\psi|^2 \right)
+ V_{\psi}(|\psi|^2)\,,
\label{eq:Ematp}\\
S_{\rm matt}^{ij}&= \frac{1}{2}\left(\rule{0mm}{0.4cm} Q^{*i}_\psi Q^j_\psi
+ Q^{*j}_\psi Q^i_\psi \right) \nonumber \\
&-\gamma^{ij}\left[ \frac{1}{2}\left( |Q_\psi|^2- 
|\Pi_\psi|^2 \right) + V_{\psi}(|\psi|^2) \right]\,, 
\label{eq:Smatp}\\
J_i^{\rm matt} &= -\frac{1}{2}\left(\rule{0mm}{0.4cm} \Pi_\psi Q_i^{*\psi}
+ \Pi^*_\psi Q_i^\psi \right)\,,
\label{eq:Jibos}
\end{align}
where we have defined, in analogy with~(\ref{eq:Q}) and
(\ref{eq:Pidef}), the variables
\begin{align}
Q_i^\psi &:= D_i\psi\,,
\label{eq:Qpsi}\\
\Pi^\psi &:= \frac{1}{\alpha}\,\frac{d\psi}{dt}
=\frac{1}{\alpha}\left(\partial_t \psi - \beta^l Q_l^\psi \right)\,.
\label{eq:Pipsi}
\end{align}
According to this, one can rewrite the KG equation~(\ref{eq:KGbos}) as
a first order PDE system like~(\ref{eq:EvQ}-\ref{eq:evPi1}).

Before finishing this Section we must emphasize the fact that boson
stars in STTs involve two distinct scalar-fields, the real-valued
non-minimally coupled field $\phi$, and the a complex-valued boson
field $\psi$.

%%%%%%%%%%%%%%%%%%%%%%%%%%%%%%%%%%%%%
%%   SCALAR GRAVITATIONAL WAVES   %%%
%%%%%%%%%%%%%%%%%%%%%%%%%%%%%%%%%%%%%

\section{Scalar gravitational waves}
\label{sec:waves}

In this section we will show that the presence of the scalar field
$\phi$ can induce the propagation of scalar (monopolar) gravitational
waves. Similar analysis have been presented before
in~\cite{PhysRevD.1.3209,PhysRevD.55.2024}. Using the weak-field
approximation, Wagoner has analyzed the properties of the source and
its radiation in the Einstein frame~\cite{PhysRevD.1.3209}. On the
other hand, Harada {\it et al.} have analyzed, in the Brans-Dicke
theory, the emission of the scalar gravitational radiation in
spherical dust collapse~\cite{PhysRevD.55.2024}. Moreover from the
detection point of view, a detailed study has been performed
in~\cite{Maggiore2000}.

Let us start by considering a linear perturbation of a flat background
metric $\eta_{ab}$ and a background scalar field $\phi_0$ such that
\begin{align}
g_{ab} &\approx \eta_{ab}+\epsilon\,\gamma _{ab}\,,
\label{eq:metricpertur}\\
\phi &\approx \phi_0+\epsilon\,\tilde\phi\,,
\label{eq:scalarpertur}
\end{align}
where $\epsilon \ll 1$ (do not confuse this $\epsilon$ with the one in
Eqs.~(\ref{phiaproxSC1}) and (\ref{phiaproxSC2})). Thus, according to
the above approximation, we have
\begin{align}
&T_{ab}\approx\,\epsilon \,\tilde{T}_{ab}\,, \\
&F(\phi)\approx\,F_{0}+\epsilon\,\tilde\phi\,F'_0\,,
\end{align}
where the subindex $0$ indicates quantities at zero order. Notice that
one must in fact have $T^{ab}_0=0$ in order to be consistent with
first order perturbations of a flat
background~\cite{salgado:2002ji}.

In the generalized Lorentz gauge $\partial^a \tilde{\gamma}_{ab} = 0$,
where $\tilde{\gamma}_{ab}$ is defined as
\begin{equation}
\tilde{\gamma}_{ab}:= \gamma _{ab} -\frac{1}{2}\,\eta _{ab}
\left( \gamma + 2\,\tilde{\phi}\,\frac{F_{0}^{\prime }}{F_{0}}\right)\,,
\label{eq:tildegamma}
\end{equation}
the field equations (\ref{eq:Einst}) and (\ref{eq:KG}) (for $V\equiv
0$) become~\cite{salgado:2002ji}
\begin{align}
&\Box _{\eta}\tilde{\gamma}_{ab} = -16\,\pi\,\frac{G_{0}}{F_{0}}\,
\tilde{T}_{ab}^{{\rm matt}}\,,
\label{eq:wavetilgamma3} \\
&\Box _{\eta }\tilde{\phi} = 4\,\pi\, \zeta\,
\frac{F_{0}^{\prime }}{F_{0}}\tilde{T}_{{\rm matt}}\,,
\label{eq:wavetilphi} 
\end{align}
where $\Box _{\eta }$ is the D'Alambertian operator in the flat
background metric, and the constant $\zeta$ is defined
as~\cite{salgado:2002ji}
\begin{align}
\zeta &:=\frac{1}{8\,\pi}\,\left(1 + \frac{3(F^{\prime}_0)^2}{16\pi F_0 G_0}
\right)^{-1}\,.
\label{eq:constantzeta}
\end{align}
Notice that the flat metric is used to raise and lower indices of
first order tensorial quantities.

We will now perform the analysis of scalar gravitational waves in a
spherical and vacuum spacetime, which is enough for the purpose of the
following discussion. In that case one can neglect the tensor modes
and assume that $\tilde{\gamma}_{ab } \equiv 0$ (note that in
spherical symmetry, quadrupole radiation is absent). Hence, according
to equation (\ref{eq:tildegamma}) we obtain
\begin{equation}
\gamma _{ab}= -\tilde{\phi}\,\eta _{ab}\,\frac{F_{0}^{\prime }}{F_{0}}\,.
\label{eq:metricandF}
\end{equation}
Therefore, the whole metric at linear order reads
\begin{equation}
g_{ab} \approx \left( 1 + \Phi\,\right)\, \eta _{ab}\,,
\end{equation}
where $\Phi:= -\tilde\phi\,F_{0}^\prime/F_{0}$ and the factor
$\epsilon$ has been re-absorbed in $\tilde{\phi}$ (note that
$\tilde{\phi}\ll \phi_0$ ({\it i.e.} $\Phi\ll 1$)).  The physical
metric then turns out to be conformally flat, $g_{ab}\approx
\Omega^2\,\eta _{a b}$, with the conformal factor $\Omega^2:= 1 +\Phi
$.

Gravitational waves are directly related with the Riemann tensor. Here
we compute that tensor using the well-known relationship between two
Riemann tensors associated with two conformal metrics (see {\em e.g.}
Eq.~(D.7) of Wald's~\cite{Wald84}). In this case, the Riemann tensor
associated with $\eta_{ab}$ vanishes and, at linear order, we have
\begin{equation}
{R_{{\rm L}\,abc}}^d =
{\delta^d}_{[a}\nabla_{b]}\,\nabla_c \Phi
- \eta^{de}\, \eta_{c[a}\,\nabla_{b]}\, \nabla_e
\Phi\,,
\label{eq:riemman}
\end{equation}
where the subindex $L$ indicates that this is valid only at the linear
approximation.  According to~(\ref{eq:riemman}), we obtain the
following components, which are directly related with the relative
(tidal) acceleration between two particles in geodesic motion,
\begin{equation}
R_{{\rm L}\,titj} =  \frac{1}{2}\,\left( -\delta_{ij}\partial_{tt}^2 \Phi
+ \nabla_i \nabla_j \Phi\right)\,.
\label{eq:riemman-proy}
\end{equation}

Assuming now that $\tilde\phi= \tilde \phi (t,r)$ is a spherically
symmetric perturbation for the scalar field, we obtain
\begin{equation}
R_{{\rm L}\,titj} = \frac{1}{2}\left[ -\delta_{ij}\partial_{tt}^2 \Phi
+ s_i s_j \partial_{rr}^2 \Phi + \frac{1}{r}\left(\delta_{ij} - s_i s_j\right)
\partial_r \Phi \right]\,.
\label{eq:riemman-proyII}
\end{equation}
where $s_i= \delta_{ij} x^j/r$ is a unit vector in the radial
direction of propagation.

On the other hand, in vacuum and in spherical symmetry, the wave
equation~(\ref{eq:wavetilphi}) implies
\begin{equation}
\partial_{tt}^2\tilde{\phi}= \partial_{rr}^2\tilde{\phi}
+ \frac{2}{r}\,\partial_r \tilde{\phi}\,.
\label{eq:relation_pt_pr}
\end{equation}
Plugging~(\ref{eq:relation_pt_pr}) into~(\ref{eq:riemman-proyII})
yields
\begin{eqnarray}
R_{{\rm L}\,titj}&=&  -\frac{1}{2}\left(\delta_{ij}
- s_i s_j\right) \partial_{tt}^2\Phi\nonumber\\  
&+& \frac{1}{2}\left(\delta_{ij} -3 s_i s_j\right) \frac{1}{r} \partial_r \Phi\,.
\label{eq:riemann-fin}
\end{eqnarray}
For outgoing radiation $\Phi(t,r)= \Psi(t-r)/r$, thus the second term
in~(\ref{eq:riemann-fin}), which involves the spatial derivative, will
be very small with respect to the first one in the ``wave zone'' and
so we can neglect it. Finally, we obtain the following expression
\begin{equation}
\label{RiemL}
R_{{\rm L}\,titj} \approx  -\frac{1}{2}\,\bot_{ij}\,\partial_{tt}^2 \Phi\,,
\end{equation}
where we have introduced the transverse projector \mbox{$\bot_{ij}=
  \delta_{ij} - s_i s_j$} in the orthogonal directions to $s_i$ and
neglected terms ${\cal O}(1/r^2)$. In the above expression, we have
considered only the massless case. When the mass term is included, a
{\it longitudinal}\/ contribution appears in Eq.~(\ref{RiemL}) ({\it
  c.f.} Ref.~\cite{Maggiore2000}).

According to~(\ref{eq:metricandF}), in the minimal coupling case,
where $F_{0}^{\prime }\equiv0$, or when one takes the value $\phi_0=0 $
asymptotically (like in the quadratic coupling $F= 1+ 8\pi\,G_{0}\xi
\phi ^2$), the scalar-gravitational waves are absent. However, taking
for $\phi_0$ the maximal value allowed by the Solar System
experiments, which constrain $\omega_{\rm BD} > 4\times 10^4$, scalar
gravitational waves are expected to develop in dynamical situations
where the NMC scalar-field is non-zero initially, like in the IS
phenomenon.  We expect the amplitude of the
$1/r$-contribution of the gravitational radiation to be of the
following order
\begin{equation}
 |R_{{\rm L}\,titj}| \approx \frac{{Q}_{scal}\, 
\omega^2}{2\,D}\,\left|\bot_{ij}\right|\,\left|F_{0}^{\prime}/F_{0}\right|\,,
\label{eq:mag-scalarwave}
\end{equation}
where $\omega$ is the frequency of the scalar wave, ${ Q}_{scal}$ is
the scalar charge (which has units of mass) and $D$ the distance to
the source.

%%%%%%%%%%%%%%%%%%%%%%%%%%%
%%%   NUMERICAL SETUP   %%%
%%%%%%%%%%%%%%%%%%%%%%%%%%%

\section{Numerical Setup}
\label{sec:numest}

In this section we present the numerical ingredients that have been
used in order to study the scalarization transition and the emission
of scalar radiation in the boson star context.

%%%%%%%%%%%%%%%%%%%%%%%
%%%   FORMULATION   %%%
%%%%%%%%%%%%%%%%%%%%%%%

\subsection{Formulation}
\label{sec:BSSN-spherical}

We perform a numerical evolution of the equations of
Sec.~\ref{sec:3+1}.  However, since the ADM equations in GR are only
weakly hyperbolic~\cite{Alcubierre08a}, we will use a formulation of
the BSSN type, which has been particularly robust in the numerical
evolution of both vacuum and matter spacetimes in
GR~\cite{Shibata95,Baumgarte:1998te}.  We adopt the particular version
of the BSSN formulation presented in~\cite{Alcubierre:2010is} that is
specially adapted to spherical symmetry.  In
Appendix~\ref{sec:Charac-BSSN} we discuss the characteristic
decomposition of this formulation for the case of a STT.

%%%%%%%%%%%%%%%%%%%%%%%%
%%%   INITIAL DATA   %%%
%%%%%%%%%%%%%%%%%%%%%%%%

\subsection{Initial Data}
\label{sec:initial-data}

We consider a single boson star in stationary equilibrium. In such a
configuration the spacetime metric is time independent, and the scalar
field $\psi(t,r)$ oscillates in time with a fixed frequency $\omega$:
\begin{equation}
\psi(t,r)=\Psi(r)\,e^{\imath\,\omega\,t}\,.
\label{eq:ansatz-phi}
\end{equation}
In order to find the initial data one must
substitute~(\ref{eq:ansatz-phi}) into the KG Eq.~(\ref{eq:KGbos}). We
now need to solve Eqs.~(\ref{eq:Einst}) and~(\ref{eq:KGbos}) in order
to obtain the frequency $\omega$, the function $\Psi(r)$, and the
metric coefficients such that, for a given amplitude of the scalar
field at the origin, $\Psi(0)$, the resulting spacetime is
static. Following~\cite{Kaup68,Palenzuela:2006wp,Palenzuela:2007dm},
we solve this problem in the polar-areal gauge, where the line element
takes the form
\begin{equation}
ds^2=-\alpha^2(r) dt^2 + A(r) dr^2 + r^2 d\Omega^2 \,,
\end{equation}
where $d\Omega^2=d\theta^2+\sin(\theta)^2\,d\phi^2$ is the usual solid
angle element.  The field equations then reduce to:
 \begin{eqnarray}
\partial_r A &=& A\,\left[ \frac{1-A}{r} \right. \nonumber \\
&+& \left. 4 \pi r \left\{ 2 A V_\psi + \frac{A \omega^2 \Psi^2}{\alpha^2}
+ Q_\psi^2 \right\} \right]\,,
\label{eq:EinsteinA} \\
\partial_r \alpha &=& \alpha \left[ \frac{A-1}{r} + \frac{\partial_rA}{2 A}
-8 \pi r A V_\psi \right] \,,\\
\partial_r \Psi &=& Q_\psi \,, \\
\partial_r Q_\psi &=& -Q_\psi \left[\frac{2}{r} + \frac{\partial_r\alpha}{\alpha}
- \frac{\partial_rA}{A} \right] \nonumber \\
&+& A \Psi \left[\partial_\psi V_\psi - \frac{\omega^2}{\alpha^2} \right] \,,
\label{eq:KGpi} 
\end{eqnarray}
where the potential $V_\psi$ is given by the
Eq.~(\ref{eq:potentialKG}).  There are several solutions of the above
system, depending on the value of the different variables at the
origin and their asymptotic behavior. In order  guarantee
regularity at the origin we impose the following boundary conditions
\begin{eqnarray}
A(0) &=& 1 \,, \label{eq:A_bcs} \\
\partial_r\alpha|_{r=0} &=& 0 \,, \\
\Psi(0) &=& \Psi_c \,, \\
Q_\psi(0) &=& 0 \,,
\end{eqnarray}
Also, for the spacetime to be asymptotically flat we must ask for
\begin{eqnarray}
A|_{\lim_{r\rightarrow \infty}} &=& 1 \,, \\
\alpha|_{\lim_{r\rightarrow \infty}} &=& 1 \,, \\
\Psi|_{\lim_{r\rightarrow \infty}} &=& 0 \,. \label{eq:psi_bcs}
\end{eqnarray}

One can consider this problem as an eigenvalue problem for $\Psi(r)$
in the sense that, for a given value of the scalar field at the origin
$\Psi_c$, the above system only admits solutions for a discrete set of
frequencies $\omega$.  We are interested in the ground state of the
boson stars, which corresponds to the configuration with no nodes in
$\Psi(r)$.

Numerical solutions of Eqs.~(\ref{eq:EinsteinA}-\ref{eq:KGpi})
satisfying the boundary conditions (\ref{eq:A_bcs}-\ref{eq:psi_bcs}),
are obtained for a given value of $\Psi_c$ by integrating the
equations from the origin outwards using a shooting
method~\cite{Press86}. This is similar to the case of neutron stars,
where the integration is also performed outwards from the origin by
giving a value of the energy-density at the center of the star. Fixing
the value of the energy density at the center of the star is
equivalent to choosing a particular stationary configuration, which in
turn determines the ADM mass as well as the total number of particles
(total baryon number in the case of neutron stars and total boson
number in the case of boson stars).

\begin{figure}[!ht]
\includegraphics[width=0.45\textwidth]{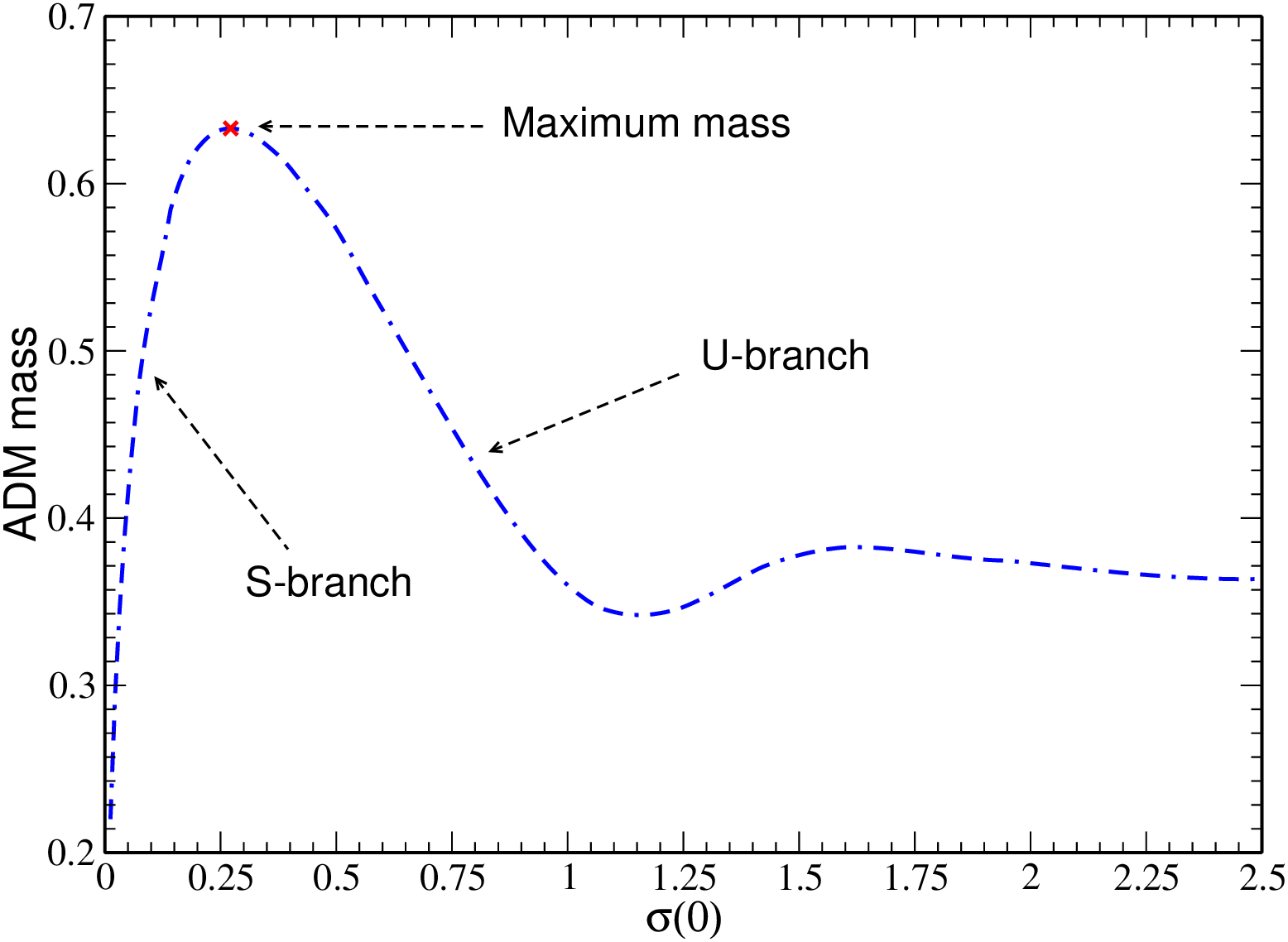}
\caption{ADM mass profile for a single boson star without
  self-interaction in GR, as a function of the central value of the
  boson field $\sigma(0)$. The maximum mass configuration, which
  corresponds to $\sigma(0)\sim 0.272$, separates the stable
  configurations (S-branch) from the unstable ones (U-branch).}
\label{fig:mass_GR.eps}
\end{figure}

Moreover, just like in neutron stars, one can construct a whole family
of stationary configurations for different energy-densities at the
origin. It can be shown that there is a specific value of the central
density which maximizes the ADM mass.  That point indicates the
threshold to configurations that are unstable to gravitational
collapse under small perturbations.  Figure~\ref{fig:mass_GR.eps}
shows the mass profile for a single boson star without
self-interaction. The maximum mass configuration, which corresponds to
a central value of the boson field given by $\sigma(0)\sim 0.272$
[where $\sigma(r):=\sqrt{4\,\pi\,G_o}\,\Psi(r)$ is a dimensionless
  boson field; {\it c.f.} Eq.~(\ref{sigma}) below], separates the
space of configurations into two branches: the stable ``S-branch'',
and the unstable ``U-branch)''. When perturbed, configurations on the
U-branch can either collapse to form black holes or disperse away
depending on value of the binding energy~\cite{Seidel90}.

In the case of STTs, there is another point of instability towards
spontaneous scalarization for a mass which is not the maximum one. In
neutron stars there is a critical baryon mass below the maximum that
marks the onset of this instability. We expect that it is the total
boson number that indicates the point of instability towards
spontaneous scalarization. However, a priori this value is difficult
to know unless one constructs a whole family of stationary
configurations and looks for solutions with a non-trivial value of the
non-minimal scalar field $\phi$. At this point is perhaps important to
remark that a maximum mass model within STTs usually corresponds to a
model with a non-trivial scalar field. That is, it corresponds to a
star that has already undergone a spontaneous scalarization process in
which some of the scalar field has been radiated away and the star has
reaches a stationary configuration. This maximum mass star is then
unstable and collapses to a black hole under a small perturbation (see
Section~\ref{sec:numres}).

%%%%%%%%%%%%%%%%%%%%%%%%%%%%%
%%%   NUMERICAL RESULTS   %%%
%%%%%%%%%%%%%%%%%%%%%%%%%%%%%

\section{Numerical results}
\label{sec:numres}

We performed numerical simulations of several single boson star
configurations. We used the code described in~\cite{Alcubierre:2010is},
which consists of a spherical reduction of the BSSN formulation,
coupled to the STT given by the Lagrangian~(\ref{eq:jordan}).  We have
taken the non-minimal coupling function~(\ref{eq:Geff}) to be of the
form
\begin{equation}
F(\phi) = 8 \pi  f(\phi) = 1 + 8 \pi G_0 \xi \phi^2 \,,
\label{eq:FNmn}
\end{equation}
with $\xi$ a positive constant. In order to have a notation consistent
with previous studies about boson
stars~\cite{Seidel90,Schunck:2003kk,Torres97}, we define the
dimensionless boson field
\begin{equation}
\label{sigma}
\sigma = \sqrt{4 \pi G_0}~\Psi \,.
\end{equation} 

All runs have been performed using a method of lines with a fourth
order Runge-Kutta integration in time, and fourth order centered
differences in space. Constraint preserving boundary have been
implemented using the algorithm described in Appendix~\ref{sec:CPBCs}.
Our typical simulations use a grid spacing of $\Delta r= 0.048$, and
we take $\Delta t = \Delta r/2$ in order to be sure that we satisfy
the Courant-Friedrich-Lewy stability condition~\cite{Alcubierre08a}.
We have also performed some simulations with grid spacings of $\Delta
r= 0.024$ and $\Delta r= 0.012$, in which case the violations on the
Hamiltonian constraint are dominated by spurious reflections from the
boundary which are, in the worst case, of order $~10^{-9}$. In all the
runs presented here the outer boundaries are located at $r_{out}=240$.

%%%%%%%%%%%%%%%%%%%%%%%%%%%%%%%%%%%%%
%%%   SPONTANEOUS SCALARIZATION   %%%
%%%%%%%%%%%%%%%%%%%%%%%%%%%%%%%%%%%%%

\subsection{Spontaneous Scalarization}

The phenomenon of scalarization in compact objects and the emission of
the monopolar gravitational waves depends strongly on the asymptotic
value of the non-minimally coupled scalar field $\phi$. If the
asymptotic value of $\phi$ is zero, which corresponds to spontaneous
scalarization, then $F^{\prime }$ tends to zero asymptotically which
means that, according to~(\ref{eq:mag-scalarwave}), there are no
monopolar gravitational waves. Nevertheless, it has been shown that in
this case the evolution of stable boson stars on STTs reaches a final
state where the scalarization ensues~\cite{Alcubierre:2010ea}. Boson
star configurations below a critical central value of the boson field
$\Psi_c^{\rm crit}$ are stable with respect to Gaussian perturbations
on the scalar field and do not lead to a SS transition: The scalar
field $\phi$ is just radiated away during the evolution.  On the other
hand, configurations above that critical value are unstable with
respect to perturbations and undergo a transition to a scalarized
state with a non-trivial scalar field $\phi$ and a non-zero scalar
charge.

%%%%%%%%%%%%%%%%%%%%%%%%%%%%%%%%%
%%%   INDUCED SCALARIZATION   %%%
%%%%%%%%%%%%%%%%%%%%%%%%%%%%%%%%%

\subsection{Induced scalarization}

In the IS phenomenon, an initially non-zero
and uniform NMC scalar field evolves naturally without the need of any
perturbation.  The reason for this is that for a non-zero background
scalar field $\phi(r)=\phi_0$ one finds $F'(\phi)|_{\phi=\phi_0} \neq
0$. The presence of the term $f'R$ in the Klein-Gordon
equation~(\ref{eq:KGo}) then implies that this initial data will not
be a stationary solution, and the field will therefore evolve away
from its initial configuration without the need for an external
perturbation.  This is in contrast with the SS case for which we have
initially $\phi(r)=\phi_0=0$, which is indeed a solution of
Eq.~(\ref{eq:KGo}), so an explicit perturbation is required to trigger
the SS phenomenon.

Since in principle any background scalar field $\phi_0$ in STTs might
disturb the constraints imposed by the Solar System experiments, the
value of $\phi_0$ must be chosen such that the corresponding
Brans-Dicke parameter $\omega_{\rm BD}$ satisfies the observational
bounds. Using the fact that the parameter $\omega_{\rm BD}$ can be
written as
\begin{equation}
\omega_{\rm BD} = \left. \frac{f}{f'^2} \right|_{\phi_0}
= \frac{1+ 8 \pi G_0 \xi \phi_0^2}{32 \pi G_0 \xi^2 \phi_0^2} \,,
\label{eq:defphi0}
\end{equation}
where $\xi$ is the NMC constant associated with the class of STTs
considered here, it turns out that
\begin{equation}
\phi_0 = \left| \frac{1}{\sqrt{8 \pi \xi  
\left( 4\omega_{\rm BD} \xi - 1 \right)}} \right| \,.
\label{eq:enforcephi0}
\end{equation}
For a given value of $\xi$, one can then enforce the constraint
$\omega_{\rm BD} \gtrsim 4.3\times 10^4$ imposed by the Cassini
probe~\cite{Bertotti2003}, and use the above expression to fix the
value for $\phi_0$ that satisfies the observational bounds.  In all
the simulations discussed below we have chosen $\phi_0$ in such a was
as to saturate the Cassine bounds, that is, the one corresponding to
$\omega_{\rm BD} = 4.3\times 10^4$ (notice that $\phi_0$ will then
depend on the parameter $\xi$).

\begin{figure}[!ht]
\includegraphics[width=0.5\textwidth]{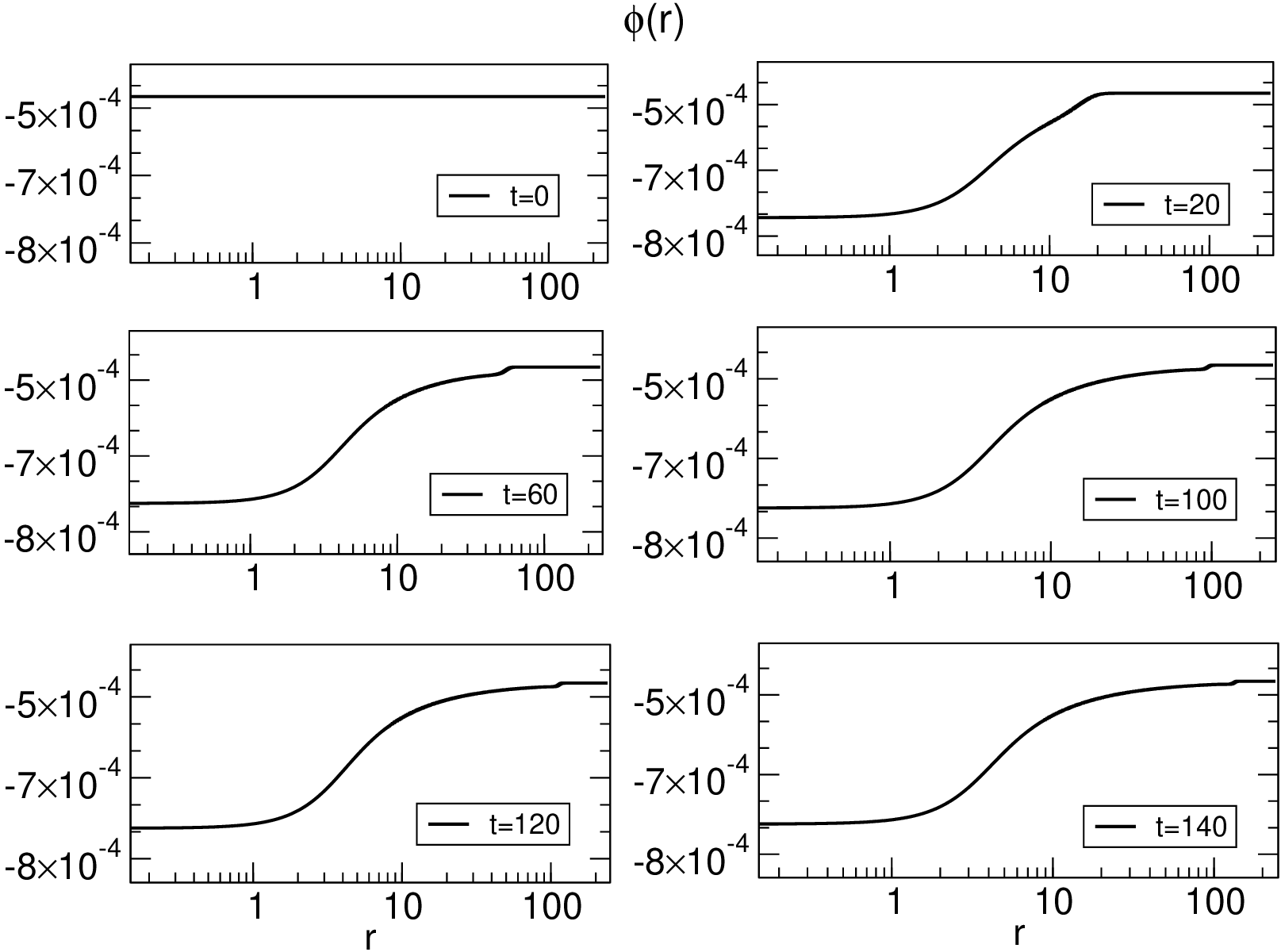}
\caption{Snapshots of the evolution of the NMC scalar field for the
  case when $\sigma(0)\sim 0.266$ initially. For this simulation we
  have taken $\xi=1$, and the NMC scalar field is initially set to
  $\phi(r)=\phi_0=-4.8\times10^{-4}$, which corresponds to the
  background value that saturates the lower bound for $\omega_{\rm
    BD}$. The system evolves until it reaches a quasi-stationary
  configuration with a non trivial scalar field $\phi$. This final
  configuration is what we refer to as ``induced
  scalarization''. Notice that asymptotically $\phi$ preserves its
  initial value $\phi_0$.}
\label{fig:evol_nm_phix1}
\end{figure}

Figure~\ref{fig:evol_nm_phix1} shows snapshots of the evolution of the
NMC field $\phi$ for a boson star with an initial central density
$\sigma(0)\sim 0.266$, which is at the threshold of the unstable
configurations. For this simulation the NMC parameter was taken to be
$\xi=1$, and the NMC scalar field is initially set to
$\phi(r)=\phi_0=-4.8\times10^{-4}$, which corresponds to the
background value that saturates the lower bound for $\omega_{\rm BD}$.
In this case a quasi-stationary configuration with a non-trivial
scalar field $\phi$ is reached at the end of the simulation.

\begin{figure}[!ht]
\includegraphics[width=0.5\textwidth]{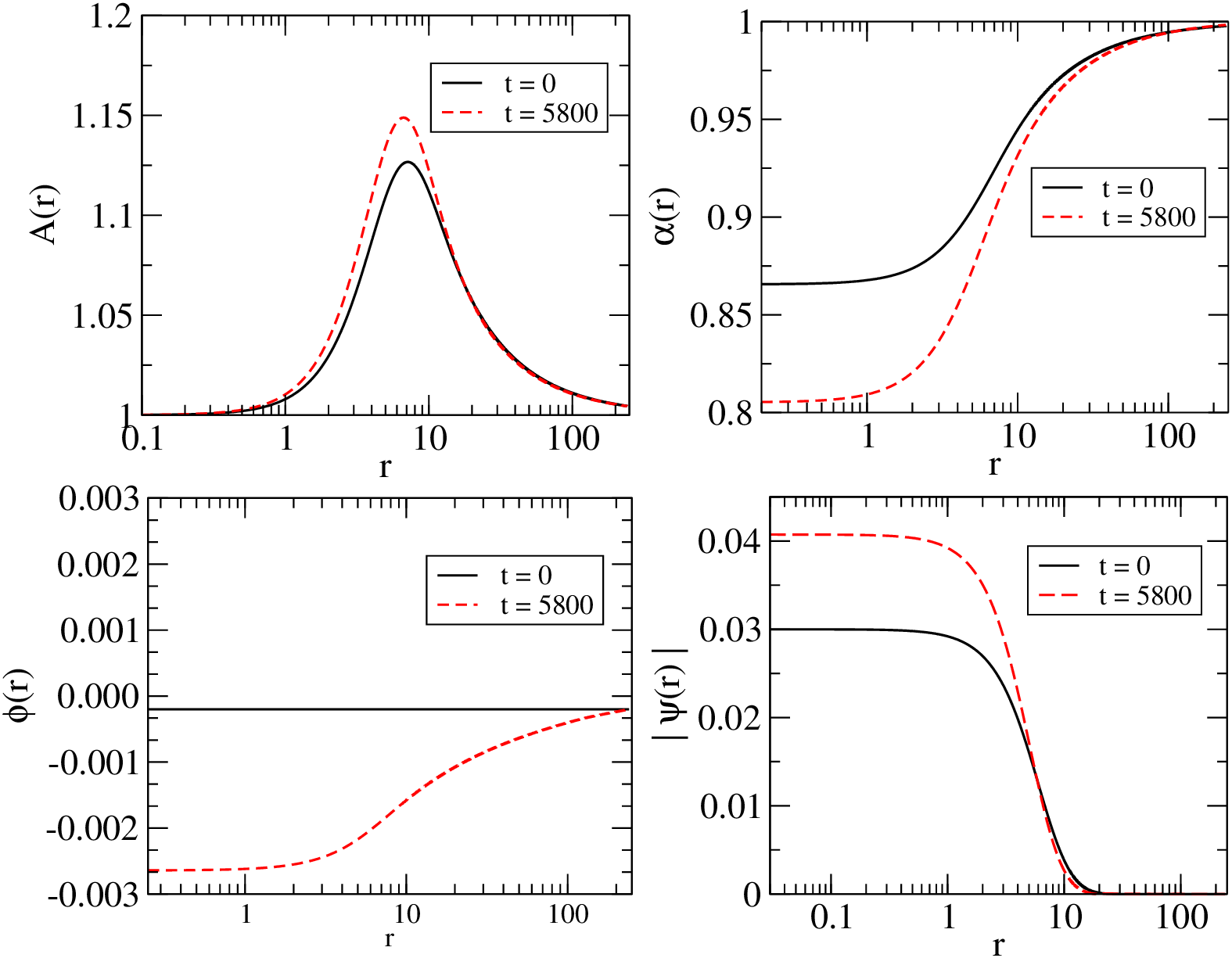}
\caption{Initial and final states of the radial metric component $A$
  (top left), lapse function $\alpha$ (top right), the NMC field
  $\phi$ (bottom left), and the norm of the complex bosonic field
  $\Psi$ (bottom right), for an initial central density $\sigma(0)\sim
  0.106$ and $\xi=500$. Just as in Fig.~\ref{fig:evol_nm_phix1}, the
  system undergoes scalarization and reaches a quasi-stationary
  configuration with a non-trivial NMC scalar field at the end of the
  evolution.}
\label{fig:snaps_metric_xi500}
\end{figure}

Figure~\ref{fig:snaps_metric_xi500} displays the initial and final
states of the radial metric component $A$, the lapse function
$\alpha$, the NMC scalar field $\phi$, and the norm of the boson field
$|\Psi|$, for a different simulation with $\xi=500$ and
\mbox{$\sigma(0) = 0.106$} initially and
$\phi_0=-9.6\times10^{-7}$. Notice that, although the parameter $\xi$
is much larger than in the previous case, the IS
only seems to affect slightly the geometry of the spacetime (confront,
for example, the initial and final values of the lapse $\alpha(0)$ in
Fig.~\ref{fig:snaps_metric_xi500}).  In order to make this statement
more precise, we first define an approximate size of our boson star
$R_{\rm star}$ as the radius of the sphere which contains the $95\%$
of the integrated mass.  The {\em compactness}\/ of the star is then
defined as $M_{\rm MS}/R_{\rm star}$, with $M_{\rm MS}$ the
Misner-Sharp mass of the system (see below).  For the simulation of
Figure~\ref{fig:snaps_metric_xi500} we find that the compactness of
the star is reduced by only $\sim 3\%$ with respect to its initial
value.

The Misner-Sharp mass $M_{\rm MS}$ used to measure the mass of the
system is defined as follows: Let $r_a$ be the areal radial coordinate,
we first define the ``Misner-Sharp mass function'' $m_{\rm MS}(r_a)$ in
terms of the radial metric component $g_{r_ar_a}(r_a)$
as~\cite{Thornburg:1998cx}
\begin{equation}
m_{\rm MS}(r_a) := \frac{r_a}{2} \; \left( 1-\frac{1}{g_{r_ar_a}(r_a)} \right) \,,
\end{equation}
with $r_a$ the areal radius.  One can show that, for asymptotically-flat
spherically-symmetric spacetimes, the Misner-Sharp mass function
coincides with the ADM mass as long as we are in vacuum.  That is,
once we are in a region outside of all the sources we should find that
$m_{\rm MS}(r_a)$ reaches a constant value $M_{\rm MS}$ such that
$M_{\rm MS} = M_{\rm ADM}$.

Several considerations are now in order.  First, even though our
initial data is in the areal gauge, during the evolution this is no
longer the case, so that one must transform back from the radial
coordinate $r$ used in the simulation to the areal radius $r_a$ in order
to calculate the mass function.  This is simple to do and we will not
go into the details here. More important, however, is the fact that
for boson stars we are never actually in vacuum since the bosonic
field $\Psi$ extends all the way to infinity.  However, one finds that
$\Psi$ decays exponentially, so that in practice for stars that are
not scalarized we very rapidly reach a region that for all practical
purposes is indeed vacuum. For scalarized stars, on the other hand, we
have to be more careful since the NMC field $\phi$ decays more slowly
(typically as $\sim 1/r$), and in the IS case it in fact reaches a
small non-zero asymptotic value $\phi_0$.  This means that the
Misner-Sharp mass function never quite reaches the constant ``vacuum''
value. Because of this, the values of $M_{\rm MS}$ we report actually
correspond to the value of the mass function $m_{\rm MS}(r)$ evaluated
at the boundary of the numerical grid. Strictly speaking this is not
the actual mass of the system, but it is a good enough approximation
for our purposes.

As a final comment one should also mention the fact that the
Misner-Sharp mass $M_{\rm MS}$ only coincides with the ADM mass for
static spacetimes.  The spacetimes considered here are dynamic, and
radiate energy during the scalarization process.  When we talk about
the initial and final mass we mean in fact the mass of the star, and
not the ``true'' ADM mass which takes into account contributions all
the way to infinity (and is therefore constant).

\vspace{5mm}

We will now try to understand the emergence of the scalarization
phenomenon on energetic grounds, where stationary scalarized configurations 
turn out to be energetically preferred over unscalarized ones.  For
instance, it has been shown in the context of neutron
stars~\cite{Damour92,Salgado98} that, beyond some critical
baryon-mass, a stationary configuration with $\phi \neq 0$ which
maximizes the fractional binding-energy of the system, is
energetically more favorable than the corresponding configuration at
the same baryon mass with $\phi \equiv 0$ (the GR case). This critical
point depends on the details of the model, such as the value of $\xi$,
the equation of state, etc.

\begin{figure}[!ht]
\includegraphics[width=0.5\textwidth]{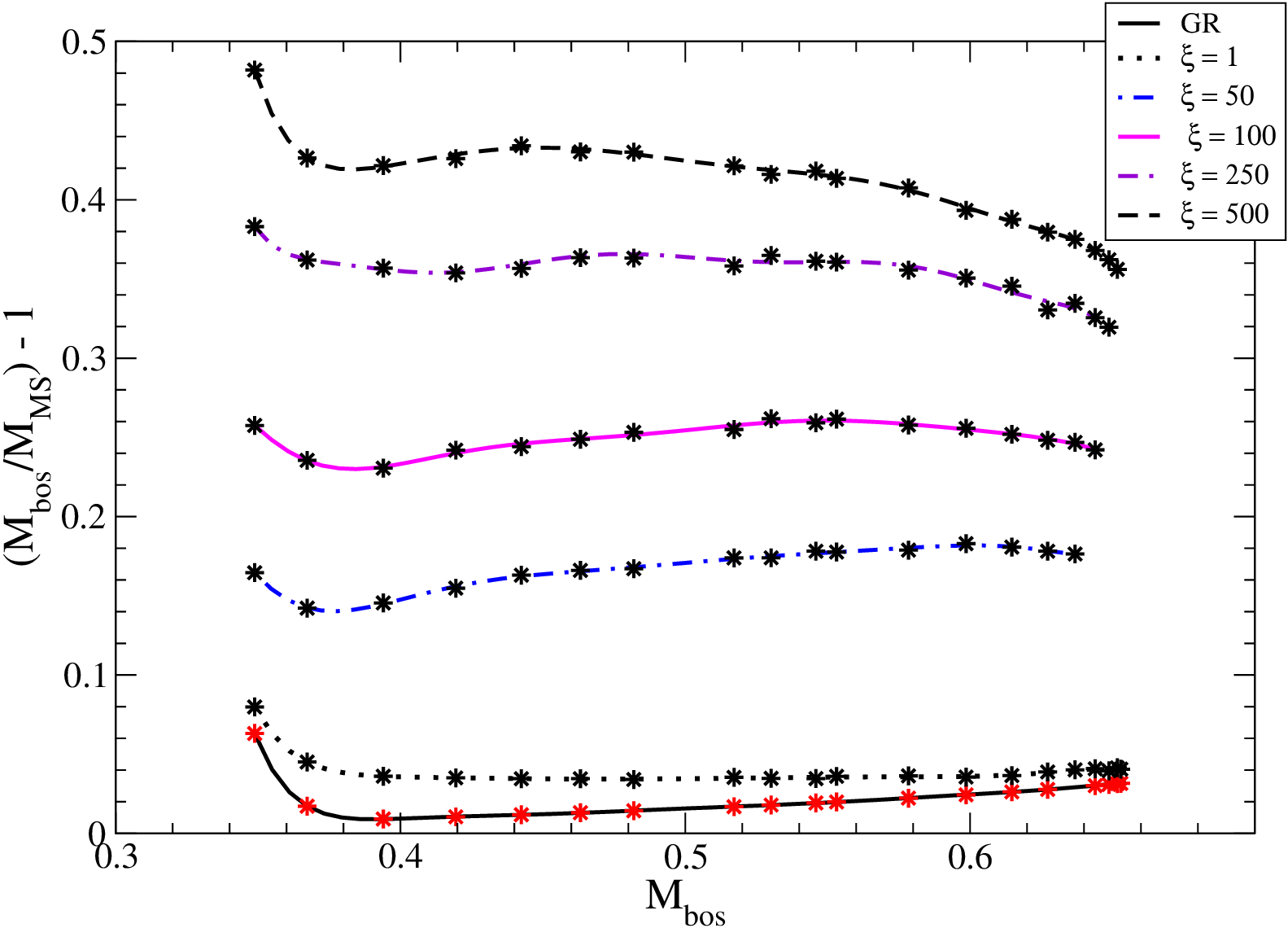}
\caption{Fractional binding energy $ (M_{\rm bos}/M_{\rm MS})-1$ as a
  function of the total boson mass $M_{\rm bos}$, for both stationary
  scalarized and unscalarized (GR case) configurations. Configurations
  with larger binding energy are energetically preferred. The points
  correspond to the actual quasi-stationary final configurations
  resulting from our simulations, while the lines are a simple fit
  obtained with polynomial regression.}
\label{fig:binding}
\end{figure}

Figure~\ref{fig:binding} shows the results of this energetic analysis
for our scalarized boson stars.  In the Figure we plot the fractional
binding-energy $ (M_{\rm bos}/M_{\rm MS})-1$ as a function of the
bosonic mass $M_{\rm bos}$ for several values of $\xi$.  Here $M_{\rm
  MS}$ is calculated as described above, while $M_{\rm bos}$ is the
mass contribution from the bosonic field alone obtained from Eq.
\eqref{eq:Tbosonmass}. From the
Figure one can see that, for a given value of $M_{\rm bos}$ and $\xi$,
the scalarized configuration is energetically preferred when compared
with an ordinary boson star with $\phi \equiv 0$ (the GR case).  It is
important to mention here that, strictly speaking, this energetic
analysis would only be valid for the SS case, where both the
scalarized and unscalarized configurations have a vanishing asymptotic
value of the NMC field $\phi$. Nevertheless, for small asymptotic
values $\phi_0$ one finds that bulk quantities are essentially the
same in the IS and SS cases, so one can still use the energetic
argument.

\begin{figure}[!ht]
\includegraphics[width=0.48\textwidth]{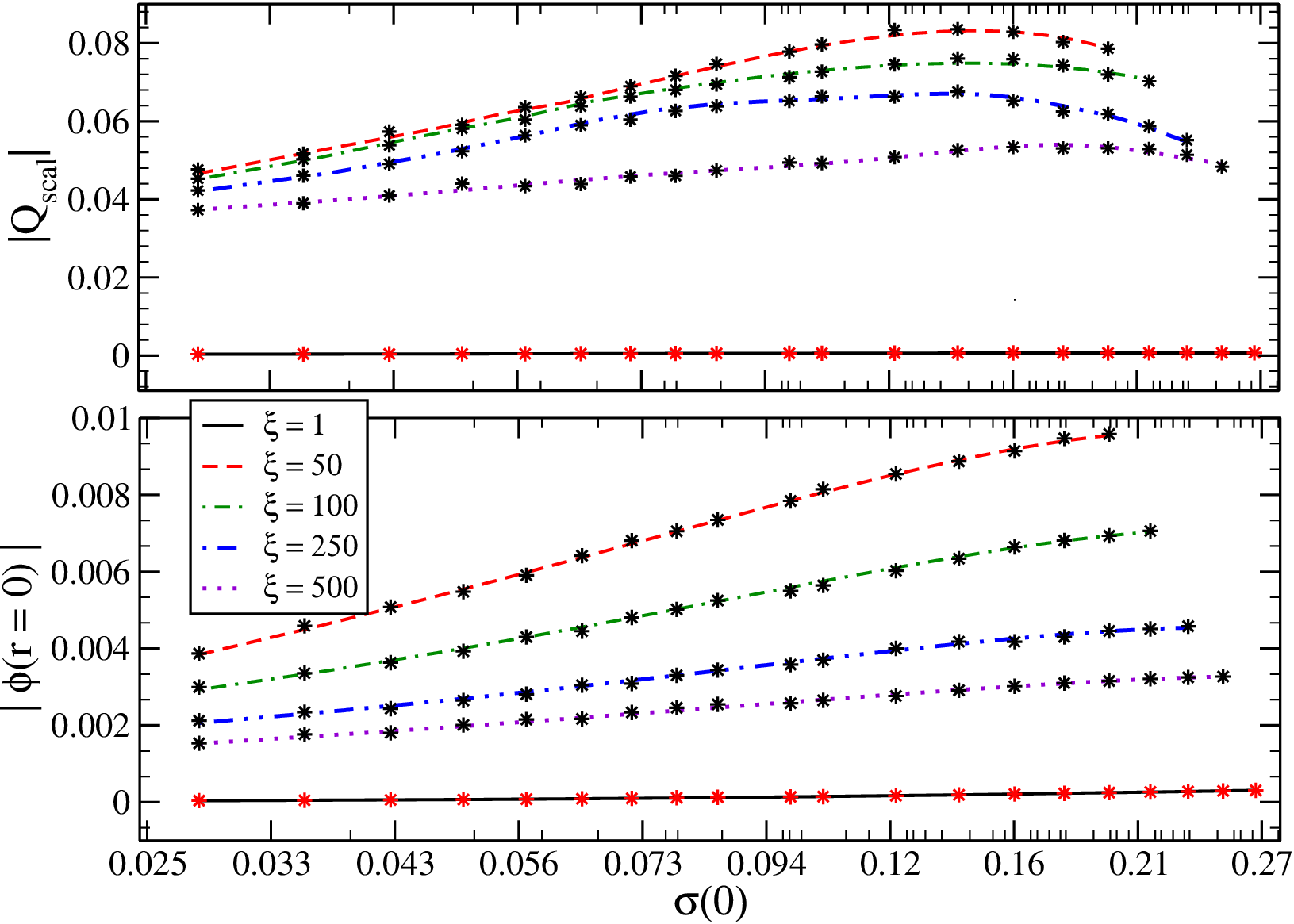}
\caption{Scalar charge (top panel), and central absolute value of the
  NMC scalar field (bottom panel), for stable boson star
  configurations as a function of the initial central density of the
  boson field $\sigma(0)$ and different values of $\xi$. Notice the
  logarithmic scale in $\sigma(0)$. }
\label{fig:nonminscalar}
\end{figure}

To corroborate the above analysis, in Figure~\ref{fig:nonminscalar} we
plot the final value of the integrated scalar charge $Q_{scal}$, and
the central absolute value of the NMC scalar field $\phi$ (which
always corresponds to its maximum), for several sets of boson star
configurations with different values of the parameter $\xi$ as a
function of the initial central density of the star $\sigma(0)$. We
have chosen initial configurations with central density in the
interval $[0.028,0.265]$, which correspond to boson stars which in GR
are stable against gravitational collapse ({\em i.e.} the S-branch
configurations of Fig.~\ref{fig:mass_GR.eps}). For these
configurations we have always assumed $\omega_{\rm BD} = 4.0\times
10^4$, which fixes the asymptotic value of the NMC field $\phi_0$.  As
expected, these stars evolve to a quasi-stationary final state with a
non-trivial scalar field $\phi$. We can conclude that the
scalarization phenomenon depends strongly on the value of the
parameter $\xi$.  For all configurations that we have analyzed we find
that the final value of the scalar charge, as well as the central
value of $\phi$, reach a maximum for $\xi \sim 50$. Notice also that
the final central value of the NMC scalar field seems to depend
logarithmically on the central density of the initial boson star
$\sigma(0)$ (in Fig.~\ref{fig:nonminscalar} we use a log scale for
$\sigma(0)$ in order to see this logarithmic dependence).

%%%%%%%%%%%%%%%%%%%%
%%%   COLLAPSE   %%%
%%%%%%%%%%%%%%%%%%%%

\subsection{Gravitational collapse and black hole formation}

Since the scalarization phenomenon modifies the compactness of the
star, one would expect that the threshold to the U-branch associated
with the scalarized stars also changes depending on the value of
$\xi$. For instance, the initial unscalarized configuration with
$\sigma(0)\sim 0.266$ and $\xi=1$, which evolved into a
quasi-stationary scalarized state as depicted in
Fig.~\ref{fig:evol_nm_phix1}, will collapse into a black hole if we
take instead $\xi\geq 10$.  This means that the critical $\sigma_{\rm
  crit}(0)$ seems to decrease when compared to the GR case in a way
that depends on the value of the parameter $\xi$. In particular, we
have found that for $\xi=50$, the last scalarized boson star
configuration that is stable against black hole formation corresponds
to a central value $\sigma(0) \sim 0.195$. This value of $\sigma(0)$
has to be contrasted with the value $\sigma(0) \sim 0.272$ for $\xi=0$
(the GR case), which is associated with the usual maximum mass
configuration.  In summary, depending on the value of $\xi$, there are
two kinds of instabilities for boson stars in STTs: On the one hand,
there is an instability that takes the star into a stable scalarized
state and, on the other hand, an instability that causes a scalarized
star to collapse to a black hole. Presumably, an unscalarized boson
star that collapses directly into a black hole may also reach a
transient (possibly very brief) state of scalarization before
collapsing.

\begin{figure}[!ht]
\includegraphics[width=0.45\textwidth]{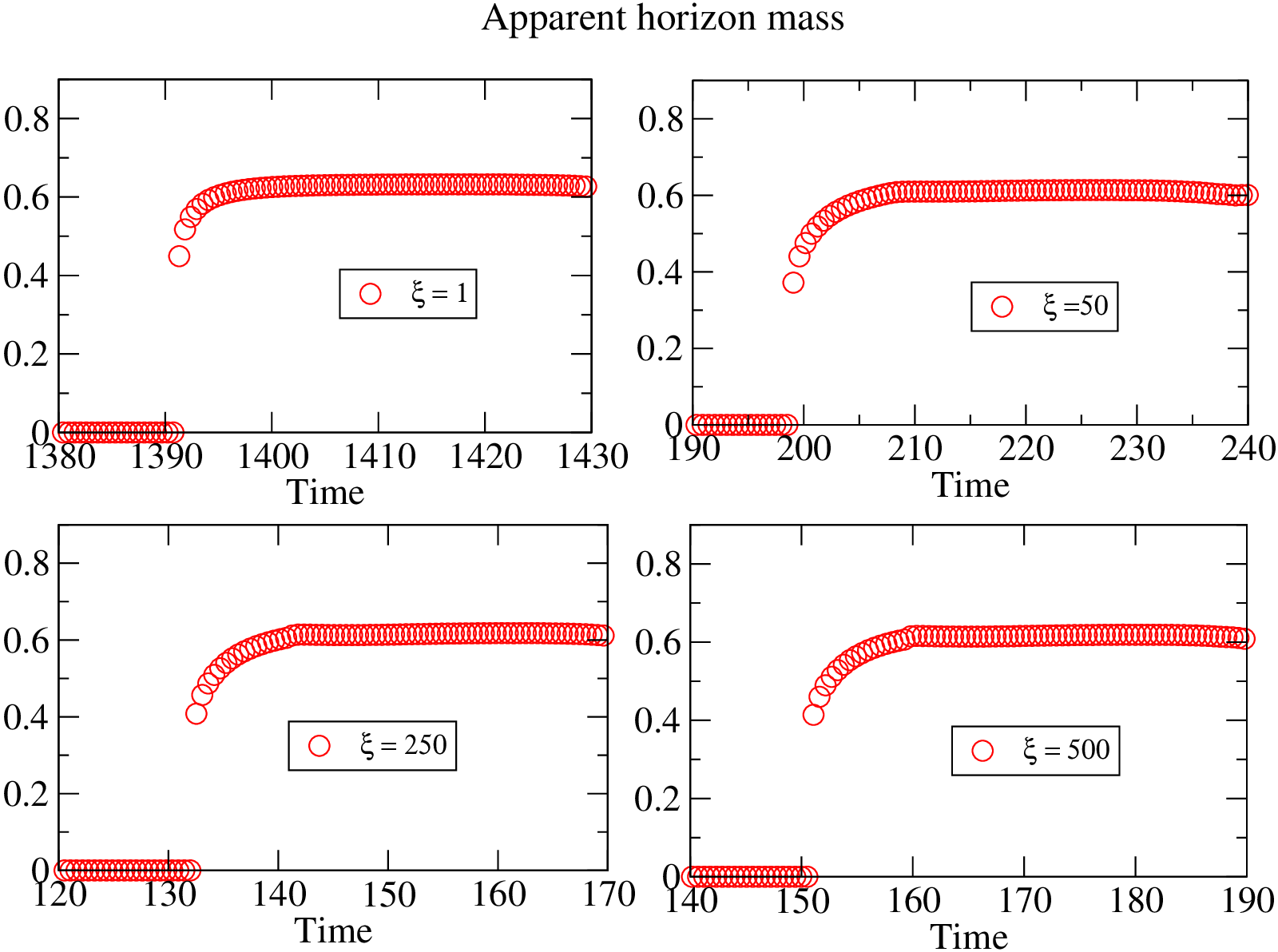}
\caption{Apparent horizon mass of the first scalarized unstable boson
  star configurations for different values of the parameter $\xi$.
  When $\xi=1$ the configuration is essentially that of GR.
  For $\xi=\{50, 250, 500\}$ the scalarized boson star
  collapses toward a black hole of the same mass but at much earlier
  times.}
\label{fig:ah_massSTT}
\end{figure}

We have studied the dynamical collapse of boson stars in STTs toward a
black hole. In order to be sure that a black hole is formed, we look
for the appearance of an apparent horizon during the simulations.  As
expected, for unstable configurations an apparent horizon appears
suddenly after some evolution. Its area then grows for some time as
more matter is accreted, until it finally settles once all the initial
matter has either fallen into the black hole or has been radiated
away.  Figure~\ref{fig:ah_massSTT} shows the apparent horizon mass
(defined in terms of its area $A$ as $M_{\rm AH} = \sqrt{A/16 \pi}$)
as a function of time, for the first unstable boson star configuration
for $\xi=\{1,50,250,500\}$ and $\sigma(0)=\{0.268,0.195, 0.231,0.249\}$. 
These values of the central density are above the threshold
given by Table \ref{table1}. The asymptotic value of the NMC field
$\phi_0$ is dependent on $\xi$ through
Eq.~(\ref{eq:enforcephi0}). Notice that as $\xi$ increases, the black
hole formation time first decreases to a certain value of $\xi$,
after which the time of collapse increases again for larger $\xi$.

\begin{figure}[!ht]
\includegraphics[width=0.5\textwidth]{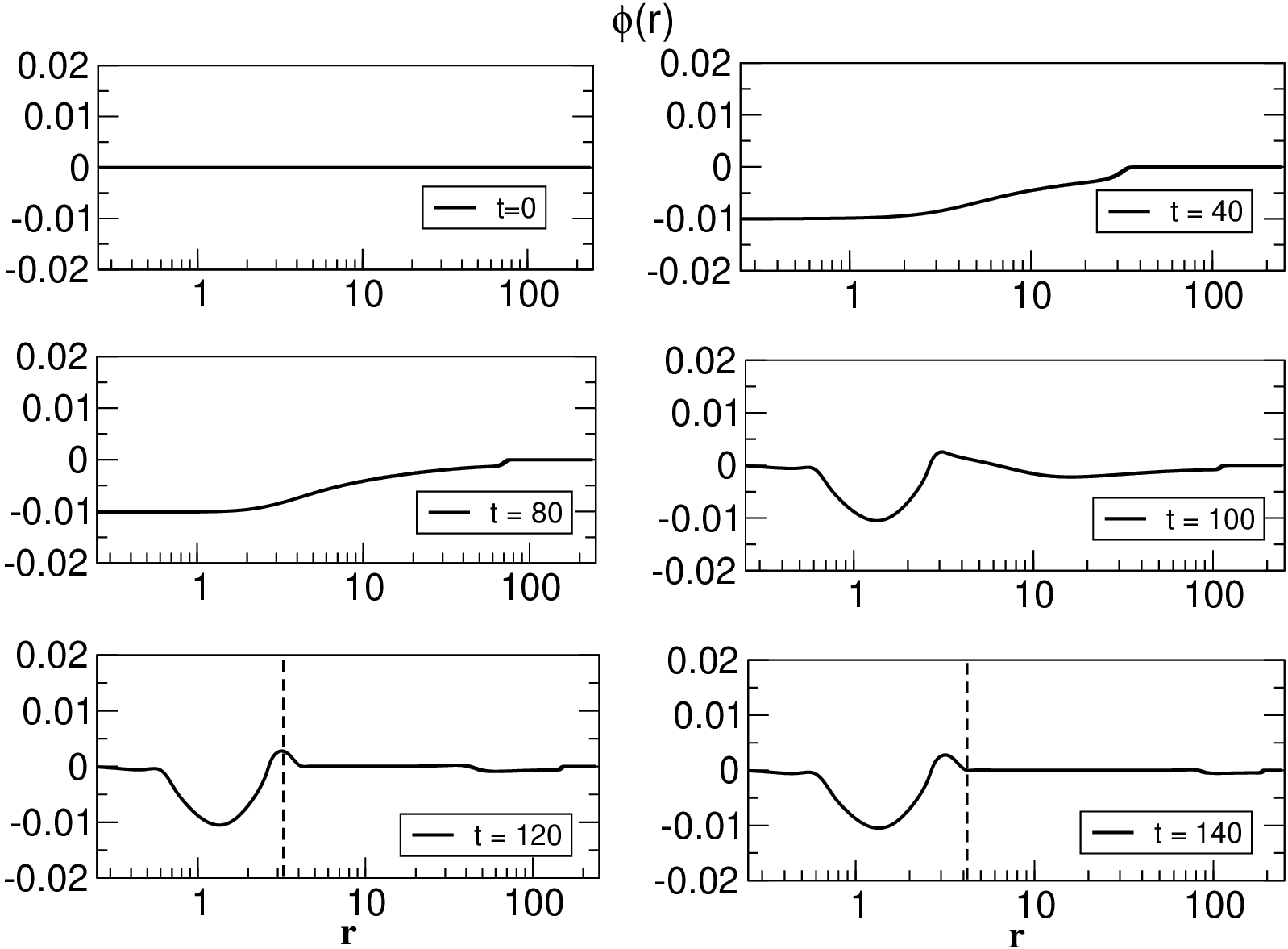}
\caption{Snapshots of the evolution of the NMC scalar field for an
  initial configuration with $\sigma(0)=0.230$ and $\xi=50$. The star
  eventually collapses into a black hole.  The same initial
  configuration with $\sigma(0)=0.230$, but taking $\xi=0$ (the GR
  case), is stable. The vertical lines in the bottom panels indicate
  the location of the apparent horizon.}
\label{fig:evol_nm_phix20}
\end{figure}

Figure~\ref{fig:evol_nm_phix20} shows some snapshots of the evolution
of the NMC scalar field for a case where $\sigma(0) = 0.23$ and
$\xi=50$, and an asymptotic value of $\phi_0=-9.25\times10^{-11}$. A
similar initial configuration with $\xi=0$ (the GR case) is stable,
while configurations with $\xi < 50$ evolve into a stable scalarized
state.

%%%%%%%%%%%%%%%%%%%%%%%%%%%%
%%%   SCALAR RADIATION   %%%
%%%%%%%%%%%%%%%%%%%%%%%%%%%%

\subsection{Scalar radiation}

Gravitational radiation can carry energy away from an isolated system,
and it also encodes important information about the physical
properties of the system itself.  In~\cite{Harada:1996wt}, Harada {\em
  et. al}\/ have performed a numerical study of the scalar
gravitational radiation emitted during an Oppenheimer-Snyder collapse
in STTs in terms of the initial parameters, such as the initial radius
and mass of the dust.

\begin{figure}[!ht]
\includegraphics[width=0.45\textwidth]{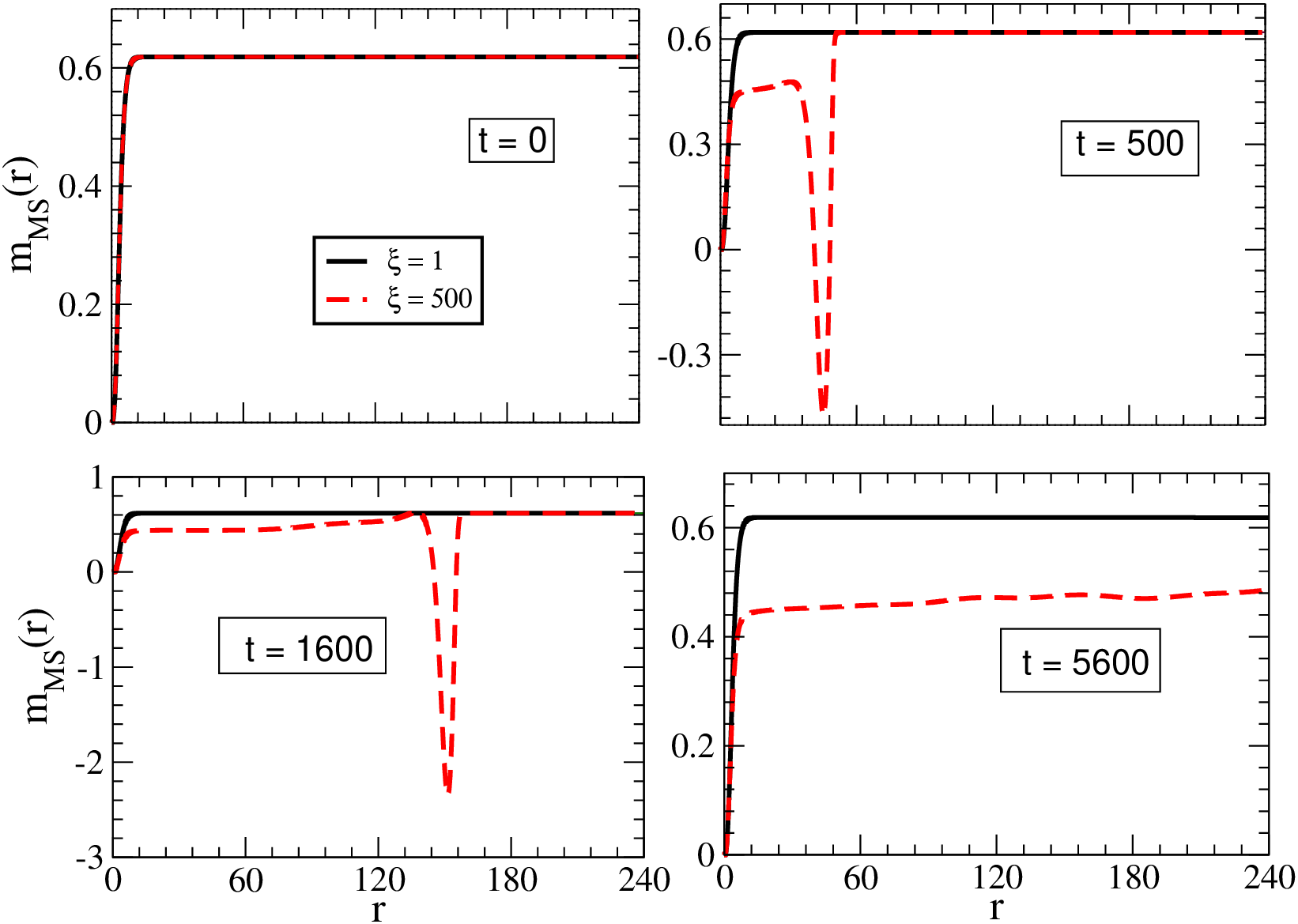}
\caption{Evolution of the Misner-Sharp mass for a stable boson star
  with central density $\sigma(0)\sim 0.141$. The variation of the
  Misner-Sharp mass with $\xi=1$ is insignificant while the variation
  of the mass with $\xi=500$ is around $20\%$ (see table
  \ref{table1}). Notice however that the peak in the plots is not
  physical since $m_{MS}$ acquires a physical meaning only in the
  limit $r\rightarrow\infty$.}
\label{fig:mass_phi}
\end{figure}

In order to study the emission of scalar gravitational radiation in
our boson star configurations, we will start by considering the
reduction of the Misner-Sharp mass during the scalarization process.
Figure~\ref{fig:mass_phi} displays the evolution of the mass function
$m_{MS}(r)$ for an initial boson star configuration with central
density $\sigma(0)\sim 0.141$, both for $\xi=1$ and $\xi=500$.  It is
clear that the configuration with $\xi=1$ reaches rapidly a
quasi-stationary state where the variation of the Misner-Sharp mass
can be ignored (it is less than $1\%$).  On the other hand, the
evolution for the configuration with $\xi=500$ is quite different.
First, one can notice a distortion moving outward for which the mass
function even becomes negative.  The position of this distortion can
be seen to coincide with an outward moving pulse of scalar field
$\phi$.  One should not worry about the fact that the mass function is
negative since, as we have said before, one can only interpret this as
a mass in the vacuum regions.  More importantly, once this pulse has
moved away, we can see a very clear reduction in the mass function. At
the end of the simulation, the final reduction in the mass is of
around $20\%$ (see table~\ref{table1}).  We should also mention the
fact that in this case the metric components and the scalar field
$\phi$ continue to oscillate at late times around a fixed
configuration (these oscillations are quite independent of the
resolution of the numerical evolution).

One could naively think that this reduction in the mass is entirely
produced by the emission of scalar {\em gravitational}
radiation. However, at this point it is difficult to separate between
the total amount of energy carried away by the scalar field $\phi$
and that carried by the scalar gravitational radiation itself.  The
latter of course arises due to the NMC between the scalar field and
the curvature, but everything is mixed in the flux of energy as given
by Eq.~(\ref{Ji}).  For instance, in the SS case with $\phi_0=0$ some
scalar field is still radiated away during the transition to the
quasi-stationary scalarized state (or during the collapse to a black
hole), even though in that case we do not expect scalar gravitational
radiation since $F'(\phi)_{\phi=0} \equiv 0$. In order to have an
unambiguous quantification of the amount of energy emitted in the
form of scalar gravitational radiation one would need to go to second
order perturbation theory and compute the equivalent of the Isaacson
energy-momentum tensor in STTs. We will leave this computation for a
future work.

\begin{widetext}
\begin{center}
\begin{table}[ht]
\begin{tabular}{l l c c c c c}
\hline \hline
$\sigma(0)\,\,\,\,\,$&$\,\,\,\xi\,/\,\phi_0\,\,$ &\,\,\,$M_{in}$ & $M_{fin}$ &
$(M_{fin}-M_{in})$ &
$E_{flux}$\\
\hline \hline
0.267 &1\,/\,$-4.8\times10^{-4}$\,\,\,   & 6.329$\times 10^{-1}$& 6.325$\times 10^{-1}\pm
1.81\times 10^{-4}$ & 4.000$\times 10^{-4}$
      &9.290$\times 10^{-6}\pm 1.41\times 10^{-8}$   \\
0.194 &50\,/\,$-9.6\times10^{-6}$\,\,\,  & 6.189$\times 10^{-1}$& 5.826$\times 10^{-1}\pm
3.55\times 10^{-3}$  & 3.630$\times 10^{-2}$
      &3.725$\times 10^{-2}\pm 3.41\times 10^{-3}$   \\
0.212 &100\,/\,$-4.8\times10^{-6}$\,\,\, & 6.251$\times 10^{-1}$& 5.611$\times 10^{-1}\pm
3.00\times 10^{-3}$ & 6.400$\times 10^{-2}$
      &6.183$\times 10^{-2}\pm 3.37\times 10^{-3}$   \\
0.230 &250\,/\,$-1.9\times10^{-6}$\,\,\, & 6.293$\times 10^{-1}$& 5.260$\times 10^{-1}\pm
3.80\times 10^{-3}$ & 1.033$\times 10^{-1}$
      &1.020$\times 10^{-1}\pm 3.42\times 10^{-3}$   \\
0.248 &500\,/\,$-9.6\times10^{-7}$\,\,\, & 6.320$\times 10^{-1}$& 5.050$\times 10^{-1}\pm
2.84\times 10^{-3}$ & 1.270$\times 10^{-1}$
      &1.271$\times 10^{-1}\pm 1.45\times 10^{-3}$   \\
\hline \hline
\end{tabular}
\caption{Initial and final values of the Misner-Sharp mass and
  integrated energy flux of the last stable boson star configuration
  in SST as measured by an Euler observer, for different values of
  $\xi$ which, according to Eq.~(\ref{eq:enforcephi0}), correspond to 
  different values of $\phi_0$.}
\label{table1}
\end{table}
\end{center}
\end{widetext}

\begin{figure}[ht]
\includegraphics[width=0.45\textwidth]{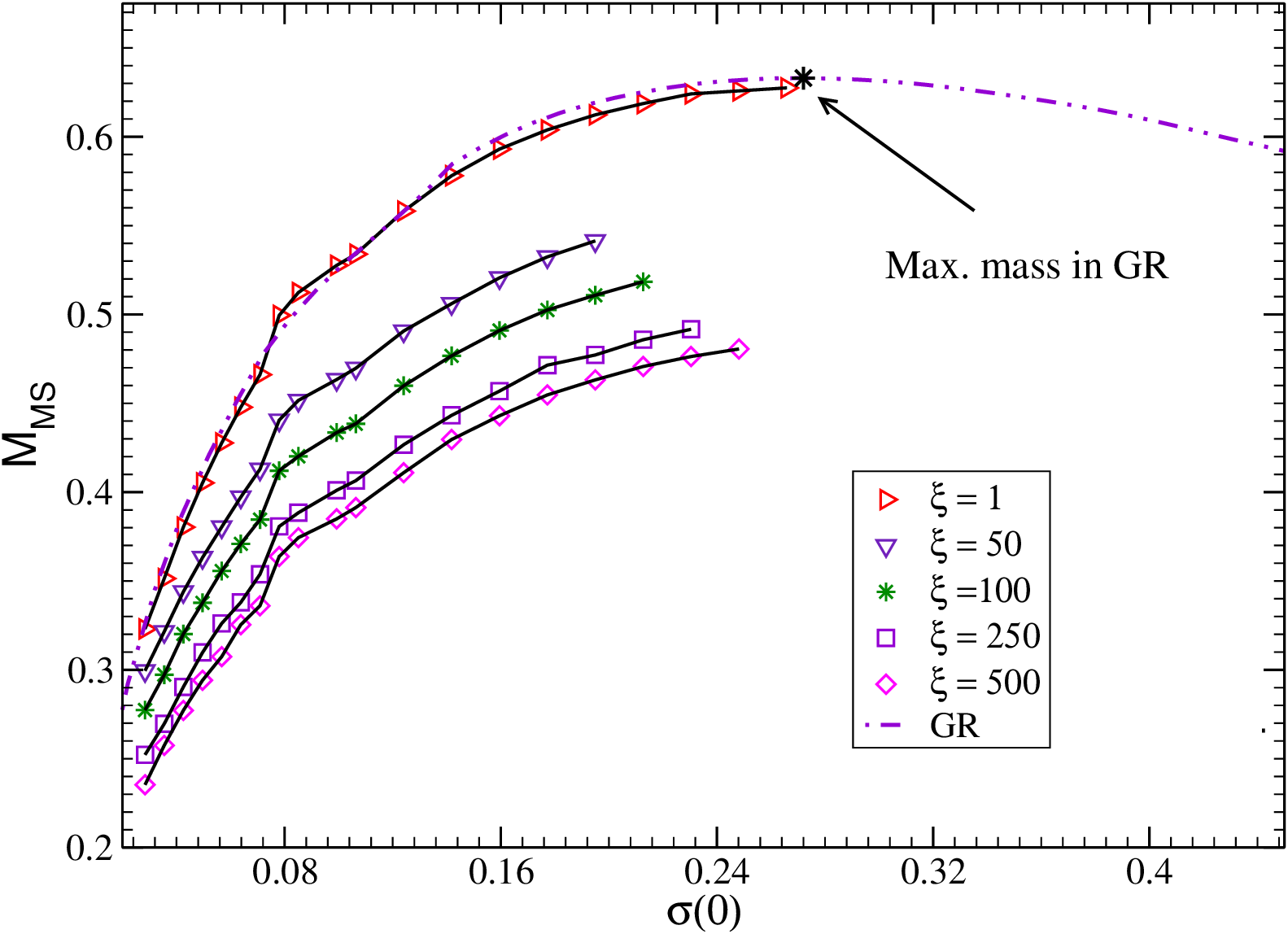}
\caption{Misner-Sharp mass for a sequence of stable stationary
  scalarized boson stars for different values of $\xi$. For reference
  we also show the GR configurations ($\xi=0$).  The last point on
  every curve corresponds to the last stable configuration found. To
  the right of that point all configurations are unstable.}
\label{fig:mass_stt}
\end{figure}

Figure~\ref{fig:mass_stt} shows the final Misner-Sharp mass for a
sequence of boson star configurations both in GR ($\xi=0$), and for
STTs with different values of $\xi$. The configurations presented here
correspond to the S-branch of a single boson star for which the system
has reached a stationary state. The stationary unstable configurations
({\em i.e.} the U-branch) would continue on the right after the last
plotted point of the curves.  We have performed several numerical
simulations using central amplitudes beyond the new S-branch in STTs,
and we have confirmed that the stars collapse into a black hole.

Table~\ref{table1} summarizes the initial and final states of the last
stable boson star configurations found for each curve of
Fig.~\ref{fig:mass_stt}.  Notice that for many of the cases studied
here, at late times the system reaches a oscillating state, with
oscillations that seem to decay very slowly, which makes it difficult
to determine the final value of the Misner-Sharp mass. So, in order to
provide an error estimation on the values reported on it, we consider
a time average between the maximum and minimum value of the ``final''
mass. We also include the value of the integrated energy flux as
measured by an Eulerian observer. Notice that the difference between
the initial and final masses agrees very well with the energy flux for
almost all the values of $\xi$, except for $\xi=1$. In this case the
oscillations of the system do not allow us to get an accurate
value for the final mass of the system.

\begin{figure}[!ht]
\includegraphics[width=0.45\textwidth]{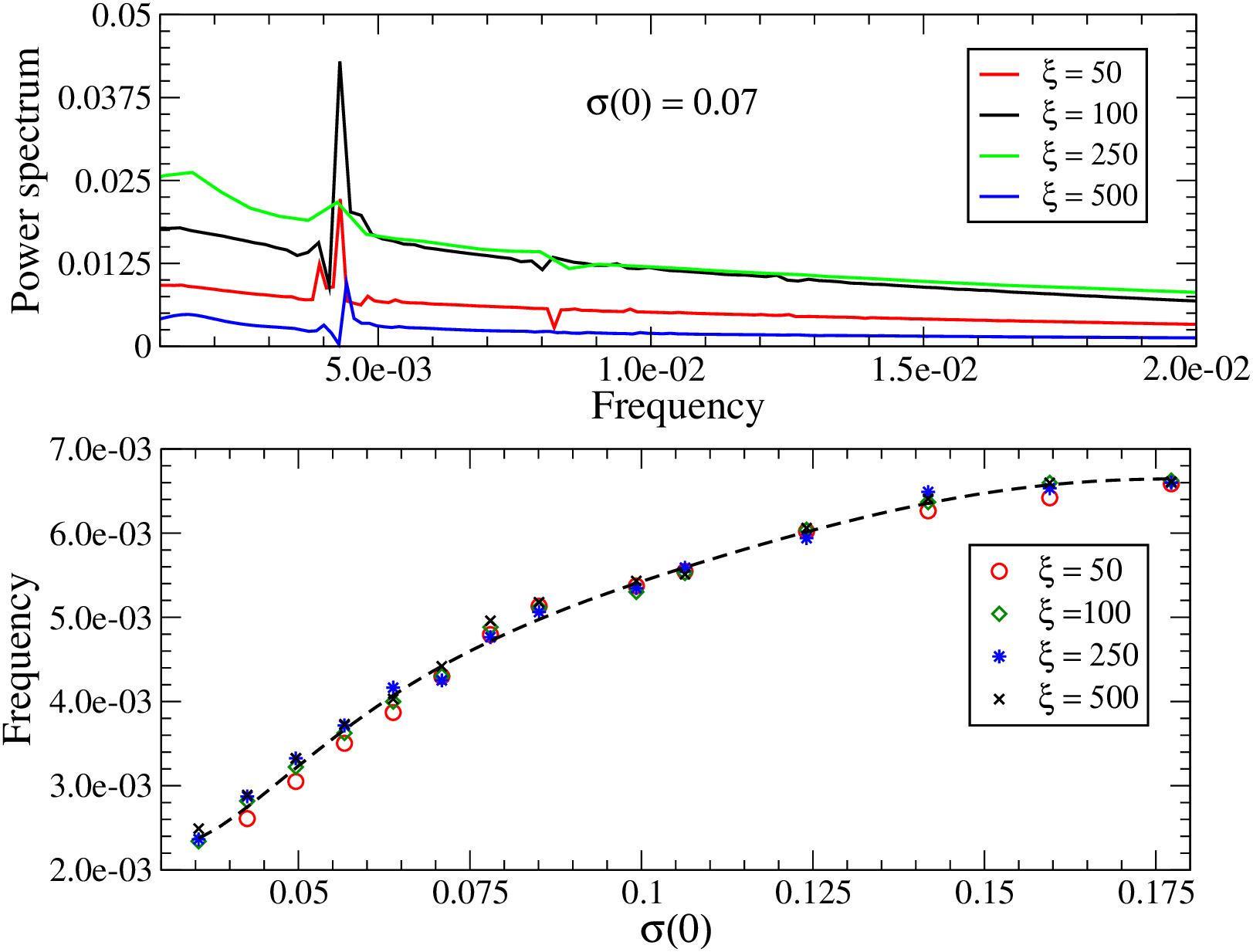}
\caption{{\em Top panel}: Fourier transform in time of the NMC scalar
  field evaluated at $r=120$ for different values
  of $\xi$ at late times (after the scalarization process).  The
  initial configuration corresponds to a boson star with $\sigma(0)=
  0.07$. {\em Bottom panel}: Main frequencies (peaks in the power
  spectrum) for different values of $\sigma (0)$.  Notice that these
  frequencies seem to be independent of the parameter $\xi$.  }
\label{fig:fftvsdenc}
\end{figure}

The evolution of the stars considered above proceeds in general as
follows: After the initial burst of scalar radiation due to the
scalarization process, the system settles down and reaches a state
with very long-lived oscillations with a characteristic frequency. In
the bottom panel of Figure~\ref{fig:fftvsdenc} we show the main
frequencies of this oscillations for the stable configurations
presented in Fig.~\ref{fig:mass_stt}.  We obtain these frequencies by
performing a discrete Fourier transform in time (at late times) of the
non-minimal scalar field at a fixed radius $r=120$, and looking for
the largest peak.  One would expect that the scalar gravitational
waves should have the same frequency~\cite{Harada:1996wt}. Notice that
the main frequency of the system seems to be independent of the
parameter $\xi$.  The top panel of the Figure shows the actual Fourier
transform (the power spectrum) for the case $\phi_0=0.07$ where one
can see that there is indeed a very clear peak.

%%%%%%%%%%%%%%%%%%%%%%%
%%%   CONCLUSIONS   %%%
%%%%%%%%%%%%%%%%%%%%%%%

\section{Conclusions}
\label{sec:discussion}

Boson stars are stable self-gravitational configurations of a complex
massive scalar field that evolves according to the Klein-Gordon
equation.  The expected mass of these type of stars typically varies
between the mass of an asteroid and a few solar masses, depending on
the mass parameter of the bosonic scalar field. some solar
masses~\cite{Colpi86}.  Though hypothetical, boson stars are simple
models that can be used to understand the corrections to general
relativity proposed by alternative theories of gravity.  Any
alternative theory of gravity has to be tested against observations,
both at the scale of the solar system and at the cosmological
scale. Scalar-tensor theories of gravity, where a (real) scalar field
is non-minimally coupled to gravity, are interesting generalizations
of general relativity that have so far not been ruled out by
observations.

In this paper we have used a scalar tensor theory of gravity to study
the evolution of distinct families of single boson stars parameterized
by the central density of the star and the parameter $\xi$ that
controls the non-minimal coupling of the theory.  We have focused on
the transition to a scalarized state using a fully relativistic code
in spherical symmetry. We have found that, just at it happens with
neutron stars, boson stars can also undergo both spontaneous and
induced scalarization, the latter case linked directly with the
emission of scalar (monopolar) gravitational waves.

Our numerical experiments show that the final magnitude of the
non-minimally coupled scalar field seems to depend logarithmically on
the central density of the boson star.  We have also found that there
is a critical value of the non-minimal coupling parameter $\xi$,
corresponding to about to $\xi \sim 50$, that maximizes the
scalarization ({\em i.e.}  the amount of final scalar charge).  On the
other hand, the maximum reduction of the initial mass of the boson
star (about $20\%$), due to energy radiated to infinity during the
scalarization process, was obtained for the maximum value of the
parameter $\xi$ considered in our evolutions ($\xi=500$).  We
summarized our results in Table~\ref{table1}. Each configuration
considered there corresponds to the critical density that separates
the stable and unstable branches of a single boson star in scalar
tensor theories for different values of
$\xi$. Figure~\ref{fig:mass_stt} shows the stable branches as a
function of this parameter, once the system has reached final
quasi-stationary state.  It is evident that, whereas a small value of
$\xi$ leads to results that are very close to those of general
relativity ($\xi=0$), larger values lead to important deviations that
in principle could be measured.

At this point it is important to stress the fact that in all the
evolutions presented here there are two parameters associated with the
specific form of the scalar tensor theory: The free parameter $\xi$
used in the expression for the non-minimal coupling function
$F(\phi)=1+8 \pi G_0 \xi \phi^2$, and the asymptotic (cosmological) value
of the scalar field $\phi_0$ that in our evolutions is chosen as the
maximum value allowed by the constraints imposed by the Cassini probe
on the effective Brans-Dicke parameter [given by
Eq.~(\ref{eq:defphi0})].  This ensures that all our results satisfy
the bounds imposed by the Solar System experiments.

Finally, we have taken a first step in trying to characterize the
monopolar gravitational waves emitted during the scalarization
process. One would expect that the magnitude of this monopolar
radiation will be proportional to the product of the scalar charge and
the square magnitude of the frequency of the radiation. We have found
that this frequency seems to be independent of the coupling parameter
$\xi$.

%%%%%%%%%%%%%%%%%%%%
%%%   APPENDIX   %%%
%%%%%%%%%%%%%%%%%%%%

\appendix

\section{Characteristic Variables for the BSSN Formulation in STTs}
\label{sec:Charac-BSSN}

Using the slicing condition~(\ref{STTBMlapse}), and assuming that the
shift vector is an {\em a priori}\/ given function of the coordinates,
Salgado {\it et. al.} have presented the characteristic decomposition
of the BSSN formulation in the STTs context~\cite{Salgado:2008xh}.
They have shown that this formulation leads to a well-posed Cauchy
problem in the Jordan frame.  In this appendix we show that the
algorithm used for dealing with the regularization of the origin, in
which one introduces a set of auxiliary variables to impose all the
required regularity
conditions~\cite{Alcubierre04A,Ruiz:2007rs,Alcubierre:2010ea}, does
not spoil those results.

Let us start by considering the spatial metric in spherical
coordinates written in the form
\begin{eqnarray}
dl^2 &=& A(t,r) dr^2 + r^2 B(t,r) d\Omega^2 \\
&=& e^{4 \chi} \left[ a(t,r) dr^2 + r^2 b(t,r) d\Omega^2 \right] \,,
\label{eq:conformal_metric}
\end{eqnarray}
where $e^{\chi}$ is the conformal factor and $d\Omega^2$ the standard
solid angle element.  In order to construct a fully first order and
regularized BSSN system, one first needs to introduce the following
set of variables
\begin{eqnarray}
d_\alpha = \p \textrm{ln} \alpha \,,\quad \quad
d_a = \frac{1}{2} \, \p \textrm{ln} a \,, \\
d_\lambda = \p \lambda \,, \quad \quad \quad\quad
\Upsilon = \p \chi \,,\\
\hat\Delta^r = \frac{1}{a} \left( \frac{\partial_r a}{2 a}
- \frac{\partial_r b}{b} - 2 r\lambda \right) \,,
\end{eqnarray}
where $\lambda$ is defined below in Eq.~(\ref{lambda}).  It turns out that, with the
above variables, the {\em principal}\/ part of the BSSN formulation in
the STTs context, coupled with the slicing
condition~(\ref{STTBMlapse}), is given by
\begin{align}
\partial_0d_\alpha&\simeq
-\alpha\,f_{\textrm{BM}}\,\left(
\p K-\frac{\theta\,f'}{f_{\textrm{BM}}\,f}\,\p \Pi\right)\,,
\label{eq:dalphamain}\\
\partial_0\Upsilon&\simeq-\frac{1}{6}\,\alpha\,\p K\,,\\
\partial_0d_a&\simeq-\frac{2}{3}\,\alpha\,r^2\,\p A_\lambda\,,\\
\partial_0 K&\simeq -\frac{\alpha\,e^{-4\,\chi}}{a}\,\left(\p d_\alpha-\frac{f'}{f}
\,\p Q_r\right)\,,
\label{eq:dKmain}\\
\partial_0A_\lambda&\simeq-\frac{\alpha\,e^{-4\,\chi}}{r^2\,a}
\,\left(\p d_\alpha+2\,\p\Upsilon
-\frac{b\,r^2}{2\,a}\,\p d_\lambda \right.\nonumber\\
&\hspace{0.5cm}\left. - a\,\p\hat\Delta^r+\frac{f'}{f}\,\p Q_r
\right)\,,\\
\partial_0d_\lambda&\simeq\frac{2\,\alpha\,a}{b}\,\p A_\lambda\,,\\
\partial_0\hat\Delta^r&\simeq\frac{2\,\alpha\,r^2}{3\,a}\,(\eta-2)\,\p A_\lambda
-\frac{2\,\alpha\,\eta}{3\,a}\,\p K \nonumber\\
&\hspace{0.5cm} + \frac{\alpha\,\eta\,f'}{a\,f}\,\p \Pi\,,\\
\partial_0Q_r&\simeq\alpha\,\p \Pi\,,
\label{eq:stt_Q}\\
\partial_0\Pi&\simeq\frac{\alpha\,e^{-4\,\chi}}{a}\,\p Q_r\,,
\label{eq:stt_Pi}
\end{align}
where we have defined $\partial_0=\partial_t+\beta^r\,\p$.

Notice that Eqs.~(\ref{eq:stt_Q}-\ref{eq:stt_Pi}) are the spherical
reduction of the Eqs. of
motion~(\ref{eq:EvQ}-\ref{eq:evPi1}). Moreover, the quantities
$\lambda$ and $A_\lambda$ are auxiliary variables needed in order to
impose regularity conditions at the origin, and which are defined
by~\cite{Alcubierre:2010ea}
\begin{equation}
\label{lambda}
\lambda:=\frac{1}{r^2}\,\left(1-a/b\right)\,,\qquad
A_\lambda=\frac{1}{r^2}\,\,\left(A_a-A_b\right)\,.
\end{equation}
Also, the equation of motion for $\hat\Delta^r$ has been modified by
adding a multiple of the momentum constraint which is controlled by
the $\eta$ parameter. In the following, we consider, for simplicity,
that $\eta=2$ (see~\cite{Alcubierre:2010ea}).

Taking into account the fact that an evolution system of the form
\begin{equation}
\partial_tu = \nu_1 \partial_rv \,, \qquad \partial_tv = \nu_2 \partial_ru\,,
\end{equation}
has the characteristic variables $w_\pm=u\mp\sqrt{(\nu_2/\nu_1)}\,v$
with characteristic speeds $\pm\sqrt{\nu_1\,\nu_2}$, it is easy to
show that the subsystem (\ref{eq:stt_Q}-\ref{eq:stt_Pi}) has the
characteristic decomposition
\begin{equation}
U^{\phi}_\pm = Q_r\pm{e^{2\,\chi}\,}\,{\sqrt{a}}\,\Pi\,,
\end{equation}
with speeds of propagation 
\begin{equation}
\mu^\phi_\pm=\beta^r\mp \alpha\,\frac{e^{-2\,\chi}}{\sqrt{a}}\,.
\end{equation}

On the other hand, the characteristic variables related with the
choice of slicing, and associated with the
subsystem~(\ref{eq:dalphamain}) and (\ref{eq:dKmain}), turn out to be
\begin{eqnarray}
U^{\textrm{gauge}}_\pm &=& d_\alpha \pm \alpha e^{2 \chi} \sqrt{a f_{\textrm{BM}}} \, K
+ \frac{f'}{f (1-f_{\textrm{BM}})} \nonumber \\
&\times& \{ Q_r (f_{\textrm{BM}}-\theta)
- e^{2 \chi} \sqrt{a f_{\textrm{BM}}} (1-\theta) \Pi \} , \hspace{8mm}
\label{eq:char-lapse}
\end{eqnarray}
with characteristic speeds
\begin{equation}
\mu^{\textrm{gauge}}_\pm = \beta^r \mp \alpha \,
\frac{\sqrt{f_{\textrm{BM}}} \, e^{-2 \chi}}{\sqrt{a}} \,.
\end{equation}

Furthermore, the following combinations
\begin{eqnarray}
U^{\lambda}_\pm &=& d_\lambda \pm
\frac{f' \left\{ Q_r (1 - 2 f_\text{BM} + \theta)
- \sqrt{a} \, e^{2 \chi} (1-\theta) \Pi \right\}}{2 f  (1-f_\text{BM})} \nonumber \\
&-& \frac{1}{2} \sqrt{a} e^{2 \chi } \left( K - A_\lambda r^2 \right) \mp
\frac{a}{2} \left( \hat{\Delta}^r + \frac{r^2 d_\lambda}{2 a^{5/2}} \right) ,
\end{eqnarray}
provide two more characteristic variables with associated speeds
\begin{equation}
\mu_\pm = \beta^r\mp \alpha\,\frac{e^{-2\,\chi}}{\sqrt{a}}\,.
\end{equation}

Finally, the  following variables
\begin{align}
U^0_1 &= \hat\Delta^r - \frac{2 f' Q_r}{a f} \left(1-\frac{2 \theta}{3 f_{\textrm{BM}}} \right)
-\frac{4 d_\alpha}{3 a f_{\textrm{BM}}} \,, \\
U^0_2 &= - a \hat{\Delta}^r + \frac{f' Q_r (2 f' f_\text{BM} - \theta)}{f f_\text{BM}}
+ \frac{d_\alpha}{f_\text{BM}} + 2 \Upsilon \,, \\
U^0_3 &= -U_2 + \frac{1}{2} \left( 3 d_a + \frac{r^2 d_\lambda}{a^{3/2}} \right) \,,
\end{align}
are the remaining characteristic variables with speed $\beta^r$.

%%%%%%%%%%%%%%%%%%
%%%%%  CPBC's  %%%
%%%%%%%%%%%%%%%%%%

\section{Constraint preserving boundary conditions}
\label{sec:CPBCs}

To reduce the influence of the numerical boundary on the dynamics of
the system we have considered constraint preserving boundary
conditions (CPBC), in the sense that small violations of the
constraints introduced by spurious reflections from the boundary
converge away with the resolution.  This is not only motivated by the
requirement of having a well-posed initial boundary problem, but
mainly because for long evolutions such as those considered here,
constraint violating modes may propagate inside the domain
contaminating the interior solution, which could lead to an incorrect
physical interpretation~\cite{Ruiz:2010qj}.

In order to impose our CPBCs we introduce information from the
characteristic variables at the boundary through the following
algorithm.

\begin{enumerate}

\item We construct numerically the characteristic outgoing modes $U_+$
  using the dynamical variables, such as gauge, metric components and
  $\phi$, at the boundary.  For instance, one can reconstruct the
  outgoing mode $U^{\phi}_+$ at the boundary by using one-sided
  difference scheme for $\Pi$ and $Q_r$.

\item In order to reconstruct the incoming scalar and gauge fields,
  $U^\phi_-$ and $U^{\rm gauge}_-$, we assume that $\phi$ and $\alpha$
  behave at the boundary as outgoing spherical waves of the form:
\begin{equation}
u(t,r)= u_0 + \frac{1}{r} \: f(r - t)\,,
\end{equation}
with $u_0$ their corresponding asymptotic values. By taking time and
space derivatives one can then reconstruct the corresponding incoming
characteristic fields.  For example, for $U^\phi_-$ we find (assuming
that $\phi=0$ asymptotically):
\begin{equation}
U^{\phi}_- = - \frac{e^{-2 \chi}}{\sqrt{a} r} \phi \,.
\end{equation}
It is interesting to notice that the incoming fields are {\em not}
zero, as one could naively expect (just setting them to zero
introduces quite large reflections).

\item This leaves us with the incoming field $U^\lambda_-$.  This is
  where one can impose the constraints, as one can show that the
  spatial derivative $\partial_r U^\lambda_-$ can be written as a
  combination of the constraints plus a term that contains no
  derivatives of the dynamical variables.  Asking then for the
  constraints to vanish at the boundary allows us to evaluate
  $\partial_r U^\lambda_-$ directly at the boundary. And once we have
  $\partial_r U^\lambda_-$ at the boundary we can use it, together
  with the values of $U^\lambda_-$ at the nearby points, to solve for
  $U^\lambda_-$ at the boundary by simple finite differencing.

\item Finally, we recover the dynamical variables using both the
  incoming and outgoing modes. Notice that it is not necessary to
  recover all the variables, since many of them are in fact not
  independent.

\end{enumerate}

\begin{figure}[!ht]
\includegraphics[width=0.45\textwidth]{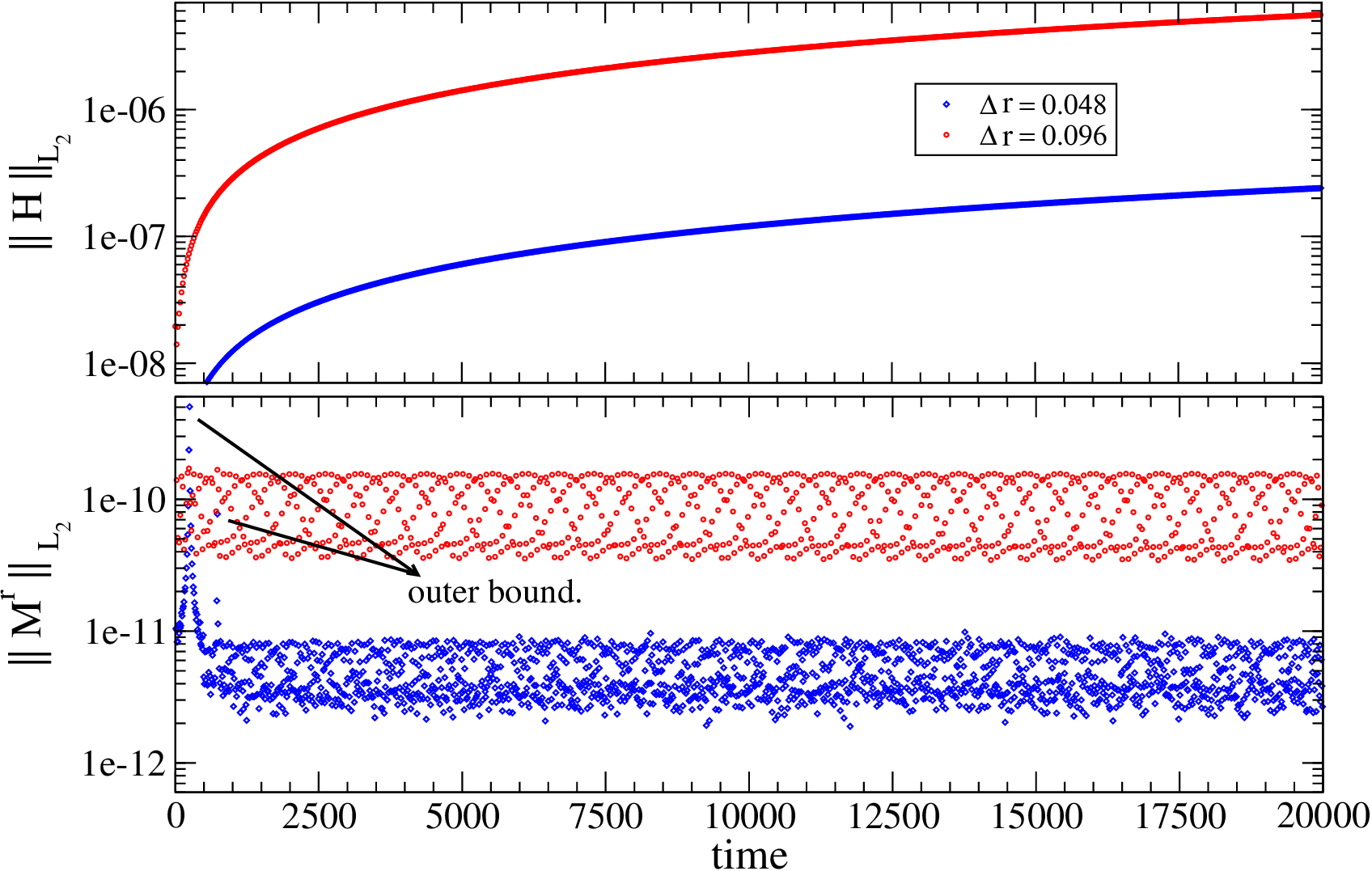}
\caption{Violation of the constraints for a boson star with
  $\sigma(0)=0.1$.  The $L_2$ norm of the Hamiltonian (top panel) and
  the momentum constraints (bottom panel) is plotted for two different
  resolutions as a function of the time.  The effect of the outer
  boundary conditions is visible only in the momentum constraint at
  times $t=240$ and $t=480$.}
\label{fig:constraints}
\end{figure}

The above algorithm allows us to impose CPBCs for the spherical
reduction of the BSSN formulation.  As an example, in
Figure~\ref{fig:constraints} we show results from an evolution of a
single boson star using forth order spatial differences and a forth
order Runge-Kutta for the integration in time. The outer boundary is
locates at $r_{out}=240$. The figure shows the $L_2$ norm of the
violation of the Hamiltonian and momentum constraints for a boson star
with $\sigma(0)=0.1$, and for two different resolutions.  Notice that
while the light crossing time of the numerical grid is $t \sim 240$,
the scheme remains fourth order convergent for much longer times.  The
boundary conditions do a spurious reflection whose magnitude is less
than $10^{-9}$. This effect is only evident in the momentum constraint
at $t=240$ and $t=480$. The magnitude of the violation of the
Hamiltonian constraint in the interior is much larger than those
reflections.

%%%%%%%%%%%%%%%%%%%%%%%%%%%
%%%   ACKNOWLEDGMENTS   %%%
%%%%%%%%%%%%%%%%%%%%%%%%%%%

\acknowledgments This work was supported in part by CONACyT grants
SEP-2004-C01-47209-F, 82787, 149945 and 132132, and by DGAPA-UNAM through
grant IN115310. JCD acknowledges DGAPA-UNAM for postdoctoral grant. MR
was supported by Spanish Ministry of Science and Innovation under
grants CSD2007-00042, CSD2009-00064 and FPA2010-16495, and 
the Conselleria d'Economia Hisenda i Innovaci\'o of the
Govern de les Illes Balears.

%%%%%%%%%%%%%%%%%%%%%%
%%%   REFERENCES   %%%
%%%%%%%%%%%%%%%%%%%%%%

\bibliographystyle{bibtex/apsrev}
\bibliography{bibtex/referencias}

%%%%%%%%%%%%%%%
%%%   END   %%%
%%%%%%%%%%%%%%%

\end{document}